\documentclass[conference]{IEEEtran}
\ifCLASSINFOpdf
\else
\fi
\usepackage{amsmath}
\usepackage{times,url,comment,cite}
\usepackage{amssymb,amsfonts,bbm}
\usepackage{graphicx}
\usepackage[caption=false,font=scriptsize]{subfig}
\usepackage{color}
\usepackage{bm}
\usepackage{xspace}
\usepackage{multirow}
\usepackage{tabularx}
\usepackage[lined,ruled,linesnumbered,noend]{algorithm2e}
\usepackage{makecell}

\usepackage{float}
\restylefloat*{figure}

\newtheorem{proposition}{Proposition}
\newtheorem{remark}{Remark}

\newcommand{\pf}{\noindent{\bf Proof:~}}
\newcommand{\qedsymb}{\hfill{\rule{2mm}{2mm}}}

\newcommand{\hv}{{\bf h}}

\newcommand{\sv}{{\bf s}}

\newcommand{\Gm}{{\bf G}}

\newcommand{\Sm}{{\bf S}}

\newcommand{\Gc}{{\cal G}}

\newcommand{\Kc}{{\cal K}}

\newcommand{\Nc}{{\cal N}}
\newcommand{\Oc}{{\cal O}}

\newcommand{\Sc}{{\cal S}}
  
\def\ben{\begin{enumerate}}
\def\beq{\begin{equation}}
\def\beqa{\begin{eqnarray}}
\def\bit{\begin{itemize}}
\def\een{\end{enumerate}}
\def\eeq{\end{equation}}
\def\eeqa{\end{eqnarray}}
\def\eit{\end{itemize}}

\def\non{\nonumber\\}

\def\argmin{\mathop{\mathrm{arg~min}}\limits}

\newcolumntype{P}[1]{>{\centering\arraybackslash}p{#1}}

\newcommand{\algobase}{\texttt{\textbf{k-merge}}\xspace}

\newcommand{\algokte}{\texttt{\textbf{kte-hide}}\xspace}
\newcommand{\anon}{$k^{\tau,\epsilon}$-anonymity\xspace}

\newcommand{\anonymized}{$k^{\tau,\epsilon}$-anonymized\xspace}
\newcommand{\anonymization}{$k^{\tau,\epsilon}$-anonymization\xspace}

\begin{document}
\title{\anon: Towards Privacy-Preserving Publishing of Spatiotemporal Trajectory Data}

\author{
	\IEEEauthorblockN{
		Marco Gramaglia\IEEEauthorrefmark{1},
		Marco Fiore\IEEEauthorrefmark{2},
		Alberto Tarable\IEEEauthorrefmark{2},
		Albert Banchs\IEEEauthorrefmark{1}%
		\vspace*{7pt}
	}
	\IEEEauthorblockA{
		\begin{tabular}[c]{P{1.15\columnwidth} P{0.7\columnwidth}}
		  \makecell{
			\IEEEauthorrefmark{1}
			IMDEA Networks Institute \& Universidad Carlos III de Madrid\\
			Avda. del Mar Mediterraneo, 22\\
			28918 Leganes (Madrid), Spain\\
			Email: name.surname@imdea.org
			}
			&
		  \makecell{
			\IEEEauthorrefmark{2}
			CNR-IEIIT\\
			Corso Duca degli Abruzzi, 24\\
			10129 Torino, Italy\\
			Email: name.surname@ieiit.cnr.it
			} \\
		\end{tabular}
	}
}


%Telephone: (800) 555--1212, Fax: (888) 555--1212}


\maketitle

\begin{abstract}
Mobile network operators can track subscribers via passive or active
monitoring of device locations. The recorded trajectories offer an
unprecedented outlook on the activities of large user populations,
which enables developing new networking solutions and services, and
scaling up studies across research disciplines.
Yet, the disclosure of individual trajectories raises significant privacy
concerns: thus, these data are often protected by restrictive non-disclosure
agreements that limit their availability and impede potential usages.
In this paper, we contribute to the development of technical solutions to
the problem of privacy-preserving publishing of spatiotemporal trajectories
of mobile subscribers.
%The solution is based on the original criterion of \anon, effective
%against record linkage and probabilistic attacks, i.e., two major
%threats to movement micro-data.
%Our solution, \algokte, implements \anon in datasets of mobile
%user trajectories. It relies on on \algobase, a low-com-plexity
%algorithm that solves a fundamental problem in movement micro-data
%anonymization, i.e., hiding spatiotemporal trajectories into one
%another through generalization.
We propose an algorithm that generalizes the data so that they
satisfy \anon, an original privacy criterion that thwarts attacks on
trajectories.
Evaluations with real-world datasets demonstrate that %\algobase and
our algorithm attains its objective while retaining a substantial level of
accuracy in the data.
Our work is a step forward in the direction of open,
privacy-preserving datasets of spatiotemporal trajectories.
\end{abstract}


% \fi
\IEEEpeerreviewmaketitle

\section{Introduction}
\label{sec:intro}

Subscriber trajectory datasets collected by network operators are logs of
timestamped, georeferenced events associated to the communication activities
of individuals. The analysis of these datasets allows inferring
\emph{fine-grained} information about the movements, habits and undertakings
of vast user populations. This has many different applications, encompassing
both business and research. For instance, trajectory data can be used to devise
novel data-driven network optimization techniques~\cite{zheng16} or support
content delivery operations at the network edge~\cite{paschos16}. They can also
be monetized via added-value services such as transport analytics~\cite{telefonica}
or location-based marketing~\cite{fluxvision}. Additionally, the relevance of
massive movement data from mobile subscribers is critical in research disciplines
such as physics, sociology or epidemiology~\cite{naboulsi16}.

The importance of trajectory data has also been recognized in the design
of future 5G networks, with a thrust towards the introduction of data
interfaces among network operators and over-the-top (OTT) providers to give
them online access to this (and other) data. OTTs can leverage such
interfaces to automatically retrieve the data and process them on the
fly, thus enabling new applications such as intelligent
transportation~\cite{asif16} or assisted-life services~\cite{czibula09}.

All these use cases stem from the disclosure of trajectory datasets to third
parties. However, the open release of such data is still largely withhold,
which hinders potential usages and applications. 
A major barrier in this sense are privacy concerns: data circulation exposes
it to re-identification attacks, and cognition of the movement patterns
of de-anonymized individuals may reveal sensitive information about them.

This calls for anonymization techniques. The common practice operators
adhere to is replacing personal identifiers (e.g., name, phone number,
IMSI) with pseudo-identifiers (i.e., random or non-reversible hash values).
Whether this is a sufficient measure is often called into question,
especially in relation to the possibility of tracking user
movements. %~\cite{narayanan14}.
What is sure is that pseudo-identifiers have been repeatedly proven not 
to protect against user trajectory uniqueness, i.e., the fact that
mobile subscribers have distinctive travel patterns that make them
univocally recognizable even in very large populations~\cite{zang11,de-montjoye13,gramaglia15}.
Uniqueness is not a privacy threat per-se, but it is a vulnerability
that can lead to re-identification.
%An example is brought forth by a recent attempt at cross-correlating
%trajectories in a mobile traffic dataset with georeferenced check-ins
%of Flickr and Twitter users: the attack could reveal the identity of
%multiple individuals with a 90\% confidence interval~\cite{cecaj14}.
Examples are brought forth by recent attempts at cross-correlating
mobile operator-collected trajectories with georeferenced check-ins
of Flickr and Twitter users~\cite{cecaj14}, with credit card records~\cite{riederer16}
or with Yelp, Google Places and Facebook metadata~\cite{mayer16}.

%%% --- EXTENSION
%In fact, more elaborate approaches to the problem of anonymizing mobile user
%trajectories exist. However, the customary strategy employed to prevent
%uniqueness in legacy relational databases, i.e., spatiotemporal generalization,
%performs poorly in this case~\cite{zang11,de-montjoye13,gramaglia15}.
%Other techniques proposed in the literature do not suit well sparse
%spatiotemporal trajectories: they only operate on
%the spatial dimension~\cite{meyerowitz09,monreale10}, are designed for regularly
%sampled (e.g., GPS) movements~\cite{yarovoy09,abul10,domingo-ferrer12},
%or risk to disrupt data utility by considerably trimming user patterns~\cite{song14}.
%Moreover, all past approaches, including recent dedicated methods~\cite{gramaglia15},
%rely on $k$-anonymity as a privacy criterion: as explained in Sec.\,\ref{sec:reqs},
%$k$-anonymity is only a partial countermeasure to attacks on spatiotemporal trajectories.
%
More dependable anonymization solutions are needed. However, the strategies
devised to date for relational databases, location-based services, or regularly
sampled (e.g., GPS) mobility do not suit the irregular sampling, time sparsity,
and long duration of trajectories collected by mobile operators.
Moreover, current privacy criteria, including $k$-anonymity and differential
privacy, do not provide sufficient protection or are impractical in this context.
See Sec.\,\ref{sec:related} for a detailed discussion.

%%In absence of a reliable technical solution to the problem of user privacy,
%%the availability of datasets of subscriber trajectories is very limited.
%%And, it only occurs under non-disclosure agreements that hinder the
%%the reproducibility and verifiability of the concerned research.
%%In most cases, NDAs limit the scope of the activitie carried out on the
%%datasets, and prevent publication of data analysis results without prior
%%verification by the relevant authorities. This is, e.g., the solution
%%adopted in the case of the mobile traffic data used in our study.

In this paper, we put forward several contributions towards
{\it privacy-preserving data publishing (PPDP)} of mobile
subscriber trajectories.
%\footnote{PPDP is defined as the development of methods for the publication
%of information that allows meaningful knowledge discovery, and yet preserves
%the privacy of monitored subjects~\cite{fung10}.}.
Our contributions are as follows: {\em (i)}~we outline attacks that are
especially relevant to datasets of spatiotemporal trajectories;
{\em (ii)}~we introduce \anon, a novel privacy criterion that effectively
copes with the most threatening attacks above; {\em (iii)}~we develop \algobase,
an algorithm that solves a fundamental problem in the anonymization
of spatiotemporal trajectories, i.e., effective generalization;
{\em (iv)}~we implement \algokte, a practical solution based on \algobase that
attains \anon in spatiotemporal trajectory data;
{\em (v)}~we evaluate our approach on real-world datasets, showing
that it achieves its objectives while retaining a substantial level of
accuracy in the anonymized data.

\section{Requirements and models}
\label{sec:reqs}

We first present the requirements of PPDP, in Sec.\,\ref{sub:ppdp}, and
formalize the specific attacker model we consider, in Sec.\,\ref{sub:att}.
We then propose a consistent privacy model, in Sec.\,\ref{sub:priv}.

%\begin{table}[tb]
%\small
%\centering
%\caption{Standard micro-data database format.}
%\vspace*{-3pt}
%\label{tab:db_std} 
%\renewcommand{\arraystretch}{1.1}
%\setlength{\tabcolsep}{4pt}
%\begin{tabular}{|l|l|l|l|l|l|l}
%\hline
%Pseudo-id & Gender & Age & ZIP & Degree & Income & \dots \\
%\hline
%\hline
%00013701 & Male & 21 & 77005 & Bachelor & 13,000 & \dots \\
%08936402 & Male & 37 & 77065 & Master's & 90,000 & \dots \\
%42330327 & Female & 60 & 89123 & High School &  46,000 & \dots \\
%\dots & \dots & \dots & \dots & \dots & \dots & \dots \\
%%\hline
%\end{tabular}
%%\vspace*{-11pt}
%\end{table}
%
%%% --- XXX extended version in infocom16-v07 and before
%
\subsection{PPDP requirements}
\label{sub:ppdp}

PPDP is defined as the development of methods for the publication
of information that allows meaningful knowledge discovery, and yet preserves
the privacy of monitored subjects~\cite{fung10}.
The requisites of PPDP are similar for all types of databases,
including our specific case, i.e., datasets of spatiotemporal
trajectories. %of mobile users.
They are as follows.%~\cite{fung10}.

\begin{itemize}
\item[1.] {\it The non-expert data publisher.} Mining of the data
is performed by the data recipient, and not by the data publisher.
The only task of the data publisher is to anonymize the data for
publication.
\item[2.] {\it Publication of data, and not of data mining results}.
The aim of PPDP is producing privacy-preserving datasets, and
not anonymized datasets of classifiers, association rules, or aggregate
statistics.
This sets PPDP apart from privacy-preserving data mining
(PPDM), where the final usage of the data is known at dataset
compilation time.
%%% --- EXTENSION
%Clearly, PPDP grants greater flexibility to end
%users than PPDM, since it does not limit the purpose the data will
%be used for. It is also a harder problem in general.
\item[3.] {\it Truthfulness at the record level}. Each record of the
published database must correspond to a real-world subject. Moreover,
all information on a subject must map to actual activities or features
of the subject.
This avoids that fictitious data introduces unpredictable biases in
the anonymized datasets.
%%% --- EXTENSION
%\item[4.] {\it The data recipient could be an attacker.} Published data
%is openly available, thus data recipients could also be attackers.
%This makes PPDP very different from the problem of database encryption,
%which assumes that only authorized and trustworthy recipients in
%possession of the required cryptographic material can access the cleartext.
\end{itemize}

Our privacy model will obey the principles above. We stress that
they impose that the privacy model must be agnostic of data usage
(points 1 and 2), and that it cannot rely on randomized, perturbed,
permuted and synthetic data (point 3). %, nor encryption (point 4).

\subsection{Attacker model}
\label{sub:att}

%\begin{figure*}[tb]
%\centering
%%\hspace*{-8pt}
%\subfloat[$\tau=1$, $n=1$]{\label{fig:att_11}
%	\includegraphics[width=0.229\textwidth]{figures/attacker-point_v2.eps}
%}
%\hfill
%\subfloat[$\tau>1$, $n=1$]{\label{fig:att_x1}
%	\includegraphics[width=0.229\textwidth]{figures/attacker-segment_v2.eps}
%}
%\hfill
%\subfloat[$\tau=1$, $n>1$]{\label{fig:att_1x}
%	\includegraphics[width=0.229\textwidth]{figures/attacker-points_v2.eps}
%}
%\hfill
%\subfloat[$\tau>1$, $n>1$]{\label{fig:att_xx}
%	\includegraphics[width=0.229\textwidth]{figures/attacker-segments_v2.eps}
%}
%%\hspace*{-5pt}
%\caption{Attacker knowledge. (a) Single location. (b) Single sequence. (c) Multiple locations.
%(d) Multiple sequences.}
%\label{fig:att}
%\vspace*{-7pt}
%\end{figure*}

%%MG EDIT
%%Unlike PPDP requirements, the attacker model is necessarily specific
%%to the type of data we consider. The model is characterized by the
%%{\it knowledge} and {\it goal} of the adversary. 
Unlike PPDP requirements, the attacker model is necessarily specific to the
type of data we consider, and it is characterized by the {\it knowledge} and
{\it goal} of the adversary. 
The former describes the information the opponent
possesses, while the latter represents his privacy-threatening objective.

\subsubsection{Attacker knowledge}
\label{sub:knowledge}

In trajectory datasets, each data record is a sequence of spatiotemporal
samples. We assume an attacker who can track a target subscriber continuously
during any amount of time $\tau$. The adversary knowledge consists then in all
spatiotemporal samples in the victim's trajectory over a continuous%
\footnote{Non-continuous tracking in the attacker model is an interesting but
very challenging open problem. A mitigative solution realisable with our model
%(which does not impose constraints on $\tau$)
is considering a $\tau$ that covers all disjoint tracking intervals.}
%\footnote{We recognize that this is a limitation of our attacker model, and
%that a more complete model would include non-continuous user tracking. Still,
%we do not impose a maximum value to $\tau$, which can possibly span the whole
%dataset duration. Thus, the attacker knowledge we consider is a superset of
%that typically assumed by attacker models in the literature.
%Non-continuous tracking in the attacker model remains in all cases a challenging
%open problem.}
time interval of duration $\tau$.
%The interval is measured in time units, determined by the
%temporal granularity of the dataset.
%Hereinafter, we denote the attacker knowledge as $\tau$ for brevity.

\subsubsection{Attacker goal}
\label{sub:goal}

Attacks against user privacy in published data can have different
objectives, and a comprehensive classification is provided in~\cite{fung10}.
Two classes of attacks are especially relevant in the context of mobile
subscriber trajectory data. Both exploit the uniqueness of movement
patterns that, as mentioned in Sec.\,\ref{sec:intro}, characterizes
trajectory data.

\begin{itemize}

\item {\it Record linkage attacks.} These attacks aim at univocally distinguishing
an individual in the database. A successful record linkage enables cross-database
correlation, which may ultimately unveil the identity of the user.
Record linkage attacks on mobile traffic data have been repeatedly
and successfully demonstrated~\cite{zang11,de-montjoye13,gramaglia15}.
As mentioned in Sec.\,\ref{sec:intro}, they have also been used for
subsequent cross-database correlations~\cite{cecaj14,riederer16,mayer16}.

\item {\it Probabilistic attacks.}
These attacks let an adversary with partial information about an individual
enlarge his knowledge on that individual by accessing the database.
They are especially relevant to spatiotemporal trajectories, as shown by
seminal works that first unveiled the anonymization issues of mobile traffic
datasets~\cite{zang11,de-montjoye13}.
Let us imagine a scenario where an adversary knows a small set of spatiotemporal
points in the trajectory of a subscriber (because, e.g., he met the target
individual there). A successful probabilistic attack would reveal the complete
movements of the subscriber to the attacker, who could then use them to infer
sensitive information about the victim, such as home/work locations, daily
routines, or visits to healthcare structures.

\end{itemize}
Our privacy model will address
%%% --- EXTENSION
%, directly or indirectly,
both classes of attacks above, led by an adversary with knowledge described
in Sec.\,\ref{sub:knowledge}.

%\begin{figure}[tb]
%\centering
%\includegraphics[width=1.0\columnwidth]{figures/attacker-taxonomy.eps}
%\caption{Attacker model taxonomy in spatiotemporal trajectory
%data.}
%% --- EXTENSION
%%Color shades indicate different levels of attack model relevance
%%to this particular type of data.
%%Contoured cells highlight the attacker
%%models we consider in this work.}
%\label{fig:tax}
%%\vspace*{-7pt}
%\end{figure}

%Combining the knowledge and goal categories above allows drawing
%a complete taxonomy of all possible attacker models, in Fig.\,\ref{fig:tax}.
%As discussed in Sec.\,\ref{sub:att-goal},
%not all combinations are meaningful to
%mobile traffic data, and we indicate with different color shades
%the relevance of each attacker model to our context.
% --- EXTENSION
%As discussed in Sec.\,\ref{sub:att-goal}, record linkage and
%probabilistic attacks are especially critical to spatiotemporal
%trajectories data, while attribute linkage may be pertinent in
%some cases.
%The figure outlines how an adversary with vast
%knowledge of an individual's movements
% --- EXTENSION
%(i.e., $\tau \sim \mathcal{T}$ or $n \sim \mathcal{N}$)
%has no interest in running
%attribute linkage or probabilistic attacks, since he already
%possesses the information these attacks aim at revealing.

\subsection{Privacy model}
\label{sub:priv}

Our privacy model is designed following the PPDP requirements
and attacker model presented before. We start by considering suitable
privacy criteria against record linkage and probabilistic attacks, in Sec.\,\ref{sub:ka}
and Sec.\,\ref{sub:uninf}, respectively. We then show how the first
criterion is in fact a specialization of the second, in Sec.\,\ref{sub:relate},
which allows us to focus on a single unifying privacy model.
Finally, we present the elementary techniques that we employ to implement
the target privacy criterion, in Sec.\,\ref{sub:tools}.
%Overall, the privacy model we propose withstands, directly or indirectly,
%the relevant attacks against subscriber privacy in mobile traffic datasets.

\subsubsection{$k$-anonymity}
\label{sub:ka}

The {\it $k$-anonymity} criterion realizes the {\it indistinguishability
principle}, by commending that each record in a database must be
indistinguishable from at least $k-1$ other records in the same
database~\cite{sweeney02}.
In our case, this maps to ensuring that each subscriber is hidden in a
crowd of $k$ users whose trajectories cannot be told apart.
The popularity of $k$-anonymity for PPDP has led to indiscriminated use
beyond its scope, and subsequent controversy on the privacy guarantees
it can provide. E.g., $k$-anonymity has been proven ineffective
againt attacks aiming at attribute linkage (including exploits of
insufficient side-information diversity),
%%% --- EXTENSION
%\footnote{Attribute linkage -- not to be confounded with the record linkage
%attacks that we consider -- includes all attacks that leverage insufficient
%diversity of the $k$-anonymized sets in side information~\cite{srivatsa12}.}
at localizing users, or at disclosing their presence and
meetings~\cite{machanavajjhala07,shokri11,srivatsa12}. % sui16.

However, $k$-anonymity remains a legitimate criterion against record linkage
attacks on any kind of database~\cite{fung10}. Therefore,
%%% --- EXTENSION
%$k$-anonymity is our reference privacy model to protect trajectory data 
this privacy model protects trajectory data
from the first type of attack in Sec.\,\ref{sub:att}, including its variations
in~\cite{zang11,de-montjoye13,gramaglia15,cecaj14,riederer16,mayer16}.

\subsubsection{\anon}
%\subsubsection{Uninformative principle and \anon}
\label{sub:uninf}

%%% --- EXTENSION
%When considering the second type of attacks in Sec.\,\ref{sub:att},
%$k$-anonymity is not a suitable criterion anymore. In fact, no privacy
%criterion proposed to date can safeguard spatiotemporal trajectory data
%from probabilistic attacks. This forces us to define an original
%criterion, as follows.
No privacy criterion proposed to date can safeguard spatiotemporal trajectory
data from the second type of attacks in Sec.\,\ref{sub:att}, i.e., probabilistic
attacks. This forces us to define an original criterion, as follows.

The pertinent principle here is the so-called
{\it uninformative principle}, i.e., ensuring that the difference
between the knowledge of the adversary before and after accessing
a database is small~\cite{machanavajjhala07}.
In our context, this principle warrants that an attacker who knows
some subset of a subscriber's movements cannot extract from the dataset
a substantially longer portion of that user's trajectory.
%Thus, the principle effectively copes with probabilistic attacks.

\begin{figure}[tb]
\centering
\includegraphics[width=0.65\columnwidth]{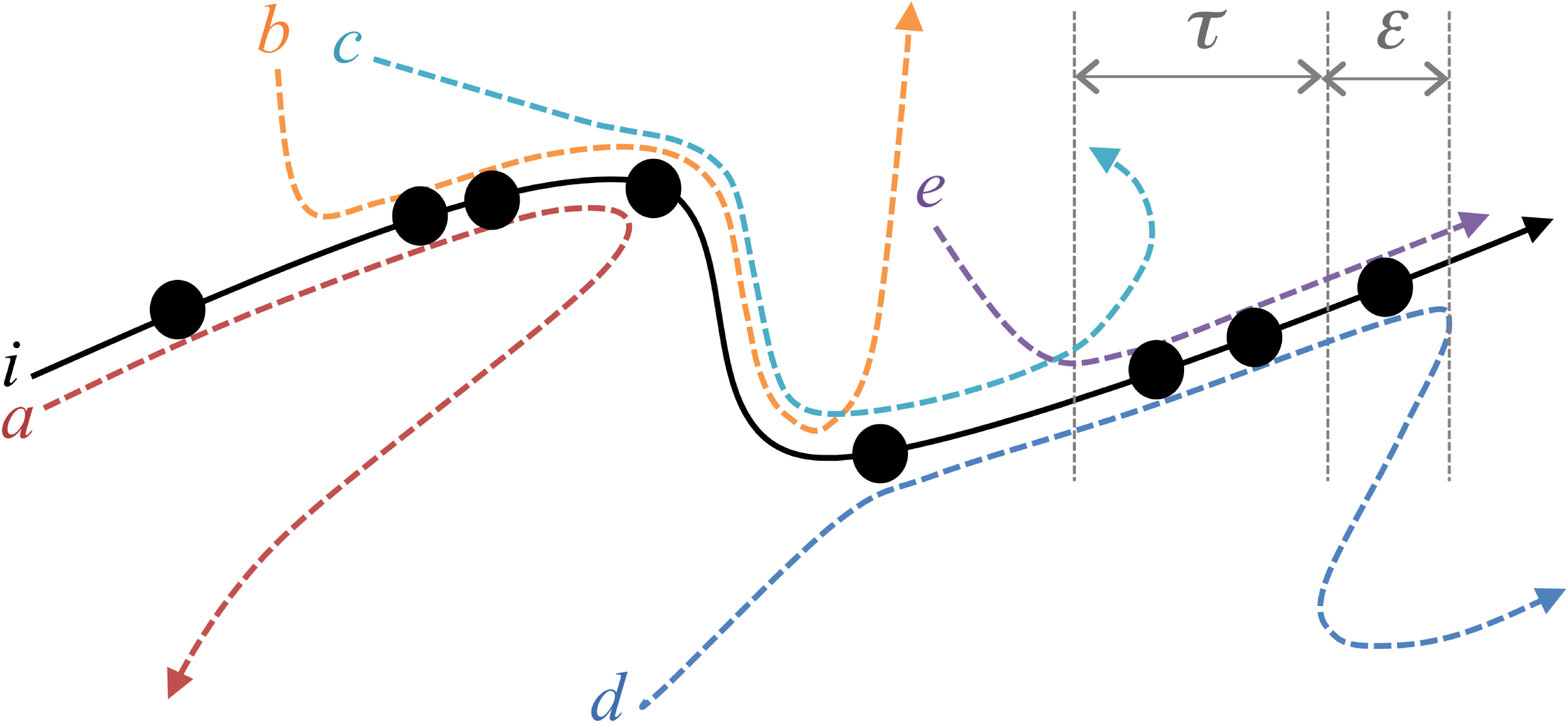}
\vspace*{-4pt}
\caption{Illustrative example of \anon of user $i$, with $k$=2.}
\label{fig:km}
\vspace*{-10pt}
\end{figure}

To attain the uninformative principle, we introduce the {\it \anon}
privacy criterion. \anon can be seen as a variation of {\it $k^m$-anonymity},
which establishes that each individual in a dataset must be indistinguishable
from at least $k-1$ other users in the same dataset, when limiting the attacker
knowledge to any set of $m$ attributes~\cite{terrovitis08}.
$k^{\tau,\epsilon}$-anonymity tailors $k^m$-anonymity to our scenario, as follows.
\begin{itemize}
\item As per Sec.\,\ref{sub:att}, the attacker knowledge
can be any continued sequence of spatiotemporal samples covering a time
interval of length at most $\tau$: thus, the $m$ parameter of $k^m$-anonymity
maps to the (variable) set of samples contained in any time period $\tau$.
During any such time period, every trajectory in the dataset must be
indistinguishable from at least other $k-1$ trajectories.
\item The maximum additional knowledge that the attacker is allowed to
learn is called {\it leakage}; it consists of the spatiotemporal samples of
the target user's trajectory contained in a time interval of duration at
most $\epsilon$, disjoint from the original $\tau$.
In order to fulfill the uninformative principle, the leakage $\epsilon$ must
be small.
\end{itemize}
The two requirements above imply alternating in time the $k-1$ trajectories
that provide anonymization.
An intuitive example
%of the rationale behind \anon
is provided in Fig.\,\ref{fig:km}.
There, the trajectory of a target user $i$ is $2^{\tau,\epsilon}$-anonymized
using those of five other subscribers. The overlapping between the trajectories
of $a$, $b$, $c$, $d$, $e$ and that of $i$ is partial and varied.
An adversary knowing a sub-trajectory of $i$ during
any time interval of duration $\tau$ always finds at least one other user with
a movement pattern that is identical to that of $i$ during that interval,
but different elsewhere. With this knowledge, the adversary cannot tell apart
$i$ from the other subscriber, and thus cannot attribute full trajectories
to one user or the other.
As this holds no matter where the knowledge interval is shifted to, the attacker
can never retrieve the complete movement patterns of $i$: this achieves the
uninformative principle.
Still, the adversary can increase its knowledge in some cases. Let us consider
the interval $\tau$ indicated in the figure: the trajectories of $i$, $d$ and $e$
are identical for some time after $\tau$, which allows associating to
$i$ the movements during $\epsilon$: the opponent learns one
additional spatiotemporal sample of $i$.

\subsubsection{Relationship between the privacy criteria}
\label{sub:relate}

It is easy to see that $k$-anonymity is a special case of \anon. As a matter
of fact, the latter criterion reduces to the former when $\tau+\epsilon$
covers the whole temporal duration of the trajectory dataset. Then,
\anon commends that each complete trajectory is indistinguishable from $k-1$
other trajectories, which is the definition of $k$-anonymity.
%
%%% --- EXTENSION
%We remark that the degenerate case above is nonsensical with respect to
%the probabilistic attacks that \anon is designed to counter. It either
%assumes that the attacker has already an almost complete knowledge of the
%trajectories ($\tau$ close to the whole dataset duration) or lets him
%acquire unbounded additional information (small $\tau$, and $\epsilon$
%spanning the rest of the dataset duration).
%Instead,
Our point here is that an anonymization solution that implements
\anon can be straightforwardly employed to attain $k$-anonymity as well,
by properly adjusting the $\tau$ and $\epsilon$ parameters.

In the light of these considerations, we address the problem of achieving
\anon in datasets of spatiotemporal trajectories of mobile subscribers.
By doing so, we develop a complete anonymization solution that is effective
against probabilistic attacks, but can also be specialized to guarantee
$k$-anonymity and counter
%the other type of relevant attacks, i.e.,
record linkage attacks.

\subsubsection{Generalization and suppression}
\label{sub:tools}

%%% --- EXTENSION
%It is extremely unlikely that a trajectory dataset fulfills the \anon criterion as
%is, even for a very small subset of subscribers in a large population~\cite{zang11,de-montjoye13}.
%We must thus enforce \anon for all users in the dataset: in other words,
%we have to tweak the spatiotemporal samples in the trajectories of individuals,
%so that the structure in Fig.\,\ref{fig:km} is respected for all of them.
%To that end, we rely on two elementary techniques, i.e.,
%{\it spatiotemporal generalization} and {\it suppression}
%of samples.
%
In order to enforce \anon for all users in the dataset, we need to
tweak the spatiotemporal samples in the trajectories of individuals,
so that the criterion in Sec.\,\ref{sub:uninf}  is respected for all of them.
To that end, we rely on two elementary techniques, i.e.,
{\it spatiotemporal generalization} and {\it suppression} of samples.

Spatiotemporal generalization reduces the precision of
trajectory samples in space and time, so as to make the samples of
two or more users indistinguishable.
Suppression removes from the trajectories those samples that are too
hard to anonymize.
Both techniques are lossy, i.e., imply some reduction of precision in
the data. Yet, unlike other approaches, these techniques conform to the
PPDP requirement of truthfulness at the record level, see Sec.\,\ref{sub:ppdp}.

\section{Achieving \anon}
\label{sec:solution}

Our goal is ensuring that an anonymized dataset of mobile subscriber
trajectories respects the uninformative principle, by implementing,
through generalization and suppression, the \anon of all subscriber
trajectories in the dataset. Clearly, we aim at doing so while minimizing
the loss of spatiotemporal granularity in the data.
% --- EXTENSION
%\begin{figure}[tb]
%\centering
%\includegraphics[width=\columnwidth]{figures/framework.eps}
%\caption{Problem decomposition.}
%\label{fig:framework}
%%\vspace*{-7pt}
%\end{figure}
% --- EXTENSION
%As the overall problem above is fairly complex to solve, we decompose
%it into a hierarchy of subproblems, depicted in Fig.\,\ref{fig:framework},
%and address them one at a time.

We start by defining the basic operation of generalizing a set of
spatiotemporal samples, and the associated cost in terms of
loss of granularity, in Sec.\,\ref{sub:gen-samp}.
We then extend both notions to (sub-)trajectories, in Sec.\,\ref{sub:gen-traj}.
Building on these definitions, we discuss in Sec.\,\ref{sub:kanon} 
the optimal spatiotemporal generalization of $k$ (sub-)trajectories.
We implement the result into \algobase,
an optimal low-complexity algorithm that generalizes (sub-)trajectories
with minimal loss of data granularity, in Sec.\,\ref{sub:kanon-impl}.
%\algobase also returns a cost information, related to the loss of granularity
%induced by the generalization: this can be used as a notion of {\it distance}
%between spatiotemporal trajectories, as it quantifies the cost of merging
%trajectories together.
%
Once able to merge (sub-)trajectories optimally, we propose an approach to
guarantee \anon of the trajectory of a single user,
%%% --- EXTENSION
%combining it with those of other subscribers,
in Sec.\,\ref{sub:kte-one}, and
% --- EXTENSION
%As our objective is not anonymizing a particular subscriber, but all those
%appearing in the same dataset, we then scale the solution to a large number
%of users. The scaling problem is not trivial,
we then scale the solution to multiple users
%%% --- EXTENSION
%-- a non-trivial passage since user trajectories need to be intertwined --
in Sec.\,\ref{sub:kte-mul}.
% --- EXTENSION
%This is a non-trivial passage since user trajectories are intertwined: e.g.,
%in the example of Fig.\,\ref{fig:km}, user $i$ is \anonymized, yet we must
%ensure the same for $a$, $b$, $c$, $d$, $e$, while respecting the constraints
%already introduced to anonymize $i$.
Finally,
%%% --- EXTENSION
%building on the different results above,
we introduce \algokte, an algorithm that ensures \anon in
spatiotemporal trajectory datasets, in Sec.\,\ref{sub:algo}.
% --- EXTENSION
%\algokte builds on the different notions
%introduced before, i.e., the \algobase algorithm and the definition
%of trajectory distance of Sec.\,\ref{sub:kanon}, and the subscriber selection
%technique of Sec.\,\ref{sub:kte-mul}.
%\algo allows ultimately anonymizing a mobile traffic dataset so that it
%fulfils the uniformative principle.
%{\color{red} Add notation table.}

\begin{figure}[tb]
\centering
\includegraphics[width=0.9\columnwidth]{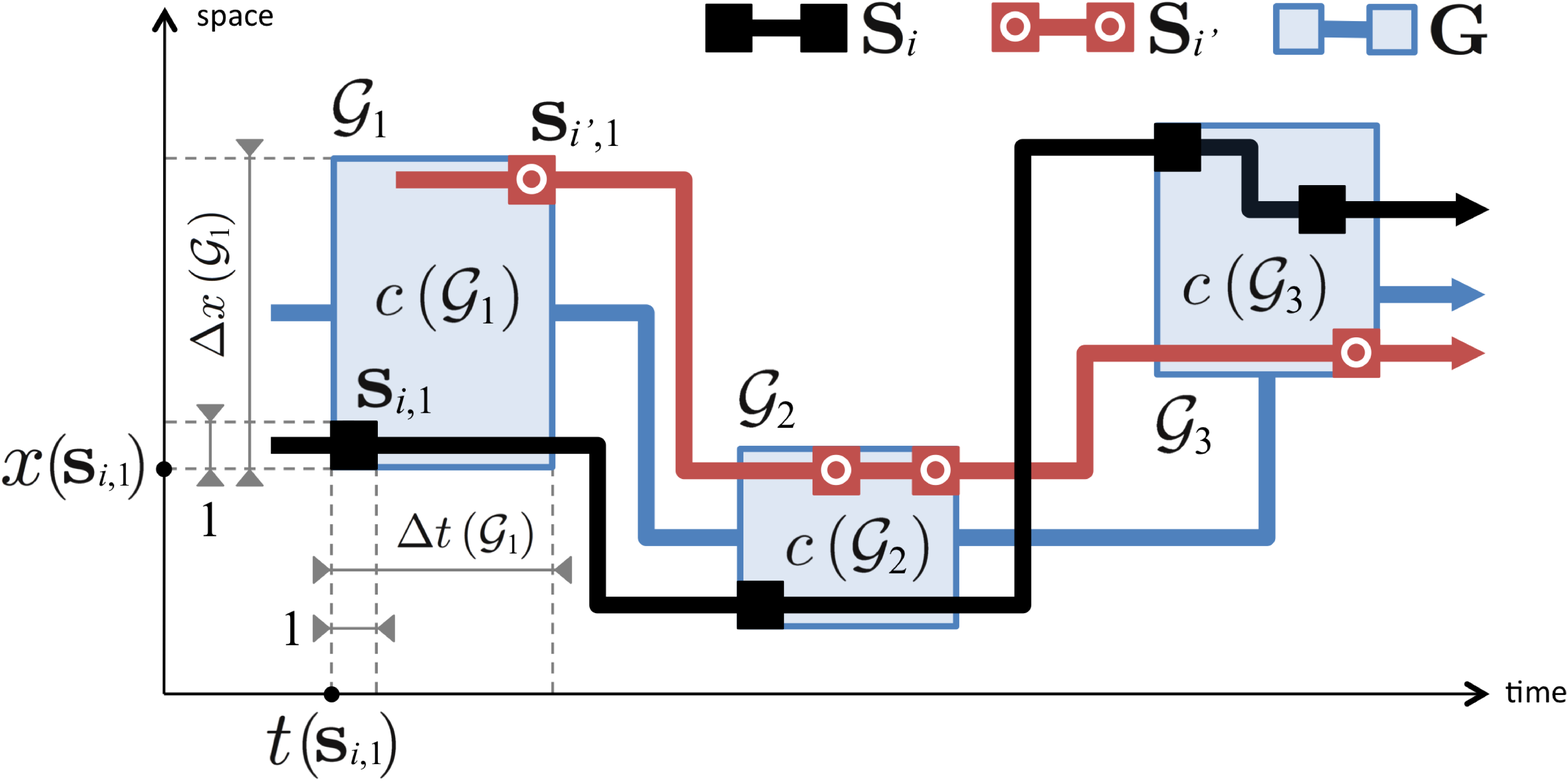}
\vspace*{-4pt}
\caption{Example of merging of trajectories $\Sm_i=\{\sv_{i,j}\}$
and $\Sm_{i'}=\{\sv_{i',j}\}$ into a generalized trajectory $\Gm=\{\Gc\}$.
For clarity, space is unidimensional.}
\label{fig:gen}
\vspace*{-10pt}
\end{figure}

\subsection{Generalization of samples}
\label{sub:gen-samp}

A (raw) \emph{sample} of a spatiotemporal trajectory represents the position of
a subscriber at a given time, and we model it with a length-3 real
vector $\sv = (t(\sv), x(\sv), y(\sv))$. Since a dataset is characterized by a finite
granularity in time and space, a sample is in fact a slot spanning some minimum
temporal and spatial intervals. The vector entries above can be regarded
as the origins of a normalized length-1 time interval and a normalized
1$\times$1 two-dimensional area%
\footnote{For instance, in our reference datasets, the sample granularity
is 1 minute in time and 100 meters in space. A raw sample spans then one
slot (i.e., 1 minute) in time and one slot (i.e., a 100$\times$100
m$^2$ area) in space. However, our discussion is general, and holds for
any precision in the data.}.

%%% --- EXTENSION
%Spatiotemporal generalization achieves anonymity by merging together two
%or more raw samples, in order to blur the information contained in each of
%them. A \emph{generalized sample} is the result of such merging and can be
%considered as a slot with a larger span.
Spatiotemporal generalization merges together two or more raw samples into
a \emph{generalized sample}, i.e., a slot with a larger span.
Mathematically, a generalized sample $\Gc$ can be represented as the set of
the merged samples.
There is a cost associated with merging samples, which is related to the span
of the corresponding generalized sample, i.e., to the loss of granularity
induced by the generalization. The cost of the operation of merging a set of samples
into the generalized sample $\Gc$ is defined as
\beq \label{eq:sample_cost}
c \left(\Gc \right) = c_t\left(\Gc\right) c_s\left( \Gc \right),
%c \left(\Gc \right) = \Delta t\left(\Gc \right) \left( \Delta x\left(\Gc \right) + \Delta y\left(\Gc \right) \right).
\eeq   
where $c_t\left(\Gc\right)$ represents the cost in the time dimension, while
$c_s\left(\Gc\right)$ is the cost in the space dimensions.

Let $\Gc_1$ and $\Gc_2$ be two disjoint generalized samples (i.e.,
$\Gc_1 \cap \Gc_2 = \emptyset$). Then, we make the following two
assumptions on the time and space merging costs:
\beq \label{eq:time_cost_prop}
c_t\left(\Gc_1 \cup \Gc_2 \right) \geq c_t\left(\Gc_1  \right) + c_t\left( \Gc_2 \right)
\eeq
\beq \label{eq:space_cost_prop}
c_s\left(\Gc_1 \cup \Gc_2 \right) \geq \max\left\{c_s\left(\Gc_1  \right) , c_s\left( \Gc_2 \right)\right\}.
\eeq

Hereafter, we use the following definitions to implement the generic
costs $c_t\left(\Gc\right)$ and $c_s\left(\Gc\right)$:
\beq \label{eq:time_cost_def}
c_t\left(\Gc \right) = \Delta t\left(\Gc \right)
\eeq
\beq \label{eq:space_cost_def}
c_s\left(\Gc \right) = \Delta x\left(\Gc \right) + \Delta y\left(\Gc \right),
\eeq
where
\beq \label{eq:sample_star}
\Delta\hspace*{-2pt}\star\hspace*{-2pt}\left(\Gc \right) = \max_{\sv \in \Gc}\hspace*{2pt} \star(\sv) - \min_{\sv \in \Gc}\hspace*{2pt} \star(\sv) + 1,
\eeq
with $\star \in \{t,x,y\}$, is the span in each dimension.
% --- EXTENSION
%The time span of $\Gc$ is then equal to
%\beq \label{eq:sample_t}
%\Delta t\left(\Gc \right) = \max_{\sv \in \Gc} t(\sv) - \min_{\sv \in \Gc} t(\sv) + 1,
%\eeq
%while its spans in the two space dimensions will be
%\beq \label{eq:sample_x}
%\Delta x\left(\Gc \right) = \max_{\sv \in \Gc} x(\sv) - \min_{\sv \in \Gc} x(\sv) + 1,
%\eeq
%\beq \label{eq:sample_y}
%\Delta y\left(\Gc \right) = \max_{\sv \in \Gc} y(\sv) - \min_{\sv \in \Gc} y(\sv) + 1.
%\eeq
%Notice that, in any dimension, the span of a generalized sample $\Gc$ is at
%least 1, and is exactly 1 if and only if all samples belonging to $\Gc$ have
%the same value in that dimension.
%The cost of the operation of merging a set of samples into the
%generalized sample $\Gc$ is then:
%\beq \label{eq:sample_cost}
%%c \left(\Gc \right) = f_t\left(\Delta t\left(\Gc \right)\right) f_{xy}\left( \Delta x\left(\Gc \right), \Delta y\left(\Gc \right) \right).
%c \left(\Gc \right) = \Delta t\left(\Gc \right) \left( \Delta x\left(\Gc \right) + \Delta y\left(\Gc \right) \right).
%\eeq

Therefore, in our implementation, $c \left(\Gc \right)$ is the
area of a rectangle with sides $\Delta t\left(\Gc \right)$ and
$\Delta x\left(\Gc \right) + \Delta y\left(\Gc \right)$.
A graphical example is provided in Fig.\,\ref{fig:gen},
where two raw samples $\sv_{i,1}$ and $\sv_{i',1}$ are merged
into a generalized sample $\Gc_1$, spanning $\Delta t(\Gc_1)$
in time and $\Delta x(\Gc_1)$ in space (portrayed as unidimensional
in the figure, for the sake of readability).

\begin{remark}
The rationale for our choice of costs is computational efficiency.
Also, summing the two space spans before multiplication allows
balancing the time and space contributions.
Finally, note that with the definition in \eqref{eq:space_cost_def},
the space merging cost assumption in \eqref{eq:space_cost_prop} is
trivially true. Instead, the definition in \eqref{eq:time_cost_def}
lets the time merging cost assumption in \eqref{eq:time_cost_prop}
hold only if the time intervals spanned by $\Gc_1$ and $\Gc_2$ are
non-overlapping.
The time coherence property that we will introduce in Sec.\,\ref{sub:gen-traj}
ensures that this is the always case.
\end{remark}

\subsection{Generalization of trajectories}
\label{sub:gen-traj}

A spatiotemporal (sub-)trajectory describes the movements of a single subscriber
during the dataset timespan.
Formally, a \emph{trajectory} is an ordered vector of samples
$\Sm = \left( \sv_1, \dots, \sv_N \right)$, where the ordering is induced by
the time coordinate, i.e., $t(\sv_i) < t(\sv_{i'})$ if and only if $i < i'$.

A \emph{generalized trajectory}, obtained by merging different trajectories,
is defined as an ordered vector of generalized samples
$\Gm =  \left( \Gc_1, \dots, \Gc_Z \right)$.
Here the ordering is more subtle, and based on the fact that the time intervals
spanned by the generalized samples are non-overlapping, a property that will be
called \emph{time coherence}. More precisely, if
$\Gc_i$ and $\Gc_{i'}$, $i < i'$, are two generalized samples of $\Gm$, then
\[
\max_{\sv \in \Gc_i} t(\sv) < \min_{\sv \in \Gc_{i'}} t(\sv).
\] 

An example of a generalized trajectory $\Gm$ merging two trajectories
$\Sm_i$ and $\Sm_{i'}$ is provided in Fig.\,\ref{fig:gen}.
$\Gm$ fulfils time coherence, as its generalized samples are
temporally disjoint.

\begin{remark}
Time coherence is a defining property of generalized trajectories in PPDP.
As a matter of fact, publishing trajectory data with time-overlapping samples
would generate semantic ambiguity and make analyses cumbersome.
%%% --- EXTENSION
%and possibly give rise to exceptions to point 3 of the PPDP requirements in Sec.\,\ref{sub:ppdp}.
\end{remark}

Analogously to the cost of merging samples, we can define a cost of merging
multiple trajectories into a generalized trajectory. We define such cost as
the sum of costs of all generalized samples belonging to it.
More precisely, if $\Gm = \left( \Gc_1, \dots, \Gc_Z \right)$, and
$c(\cdot)$ is defined as in~\eqref{eq:sample_cost}, then the cost
of $\Gm$ is given by:
\beq \label{eq:traj_cost}
C \left( \Gm \right) = \sum_{i=1}^Z c \left(\Gc_i \right).
\eeq

\begin{remark}
The cost in~\eqref{eq:traj_cost} is the overall surface covered by samples
of the generalized trajectory over the spatiotemporal plane. E.g., in
Fig.\,\ref{fig:gen}, the cost of $\Gm$ is the sum of the three areas, i.e.,
$c(\Gc_1) + c(\Gc_2) + c(\Gc_3)$.
It is thus proportional to the total loss of granularity induced by the
generalization.
%It thus inversely relates to the level of accuracy resulting from the aggregation:
%namely, a lower cost maps to a higher granularity of the generalized trajectory.
\end{remark}

\subsection{Optimal generalization of trajectories}
\label{sub:kanon}

We now formalize the problem of {\it optimal} generalization of spatiotemporal
(sub-)trajectories.
%and define a low-complexity algorithm to solve it.
Suppose that we have $k$ trajectories $\Sm_1, \dots, \Sm_k$, with $\Sm_i =
(\sv_{i,1},\dots,\sv_{i,N_i})$, $i=1,\dots,k$. The goal is a
generalized trajectory $\Gm^* = \left( \Gc^*_1, \dots, \Gc^*_Z \right)$
from $\Sm_1, \dots, \Sm_k$, which satisfies the following conditions.

\emph{i)} 
The union of all generalized samples of $\Gm^*$ must coincide with the
union of all samples of $\Sm_1, \dots, \Sm_k$, i.e.,
\[
\Gc^*_1 \cup \dots \cup \Gc^*_Z = \Sc_1 \cup \dots \cup \Sc_k \triangleq \Sc,
\]
where $\Sc_{i} = \bigcup_{j=1}^{N_i} \{\sv_{i,j}\}$. Thus, $\Gm^*$ is a
partition of the set $\Sc$ of all samples in the input trajectories: it
does not add any alien sample or discard any input sample.
%%% --- EXTENSION
%\footnote{While adding alien samples
%would violate point 3 of the PPDP requirements in Sec\,~\ref{sub:ppdp},
%discarding samples is allowed. As already said in Sec.~\ref{sub:g&s}, we resort to suppression, but in
%a preparatory stage before applying the merging algorithm.}.

\emph{ii)} Each generalized sample contains at least one sample from each of
the $k$ input trajectories $\Sm_1, \dots, \Sm_k$, i.e.,
\[
\Gc^*_i \cap \Sc_{i'} \neq \emptyset, \,\,\, i = 1,\dots,Z, \,\,\, i' = 1,\dots,k.
\]
This imposes that each
input trajectory contributes to each generalized sample of $\Gm^*$.
Otherwise, the merging could associate generalized samples to users that never
visited the generalized location at the generalized time, violating point 3
of the PPDP requirements in Sec.\,\ref{sub:ppdp}.

\emph{iii)}
The cost of the merging is minimized, i.e.,
\beq  \label{eq:opt_kanon}
\Gm^* = \argmin_{\Gm \in \Kc} C(\Gm),
\eeq
where $\Kc$ is the set of all partitions of $\Sc$ satisfying time coherence
as well as condition \emph{ii)} above, and $C(\Gm)$ is in~\eqref{eq:traj_cost}.
In Fig.\,\ref{fig:gen}, the generalized trajectory
$\Gm$ fulfils all these requirements, and is thus the optimal merge $\Gm^*$ of
$\Sm_i$ and $\Sm_{i'}$.

Solving the problem above with a brute-force search is computationally
prohibitive, since $\Kc$ has a size that grows exponentially with $|\Sc|/k$,
where $|\cdot|$ denotes cardinality. However, we can characterize $\Gm^*$ so
that it is possible to compute it with low complexity. To that end, we name
\emph{elementary} a partition $\Gm \in \Kc$ that cannot be refined to another
partition within $\Kc$. In other words, none of the generalized samples of an
elementary partition can be split into two generalized samples without
violating conditions i) and ii) above, or time coherence. Then, we have the
following proposition.

\begin{proposition}
Given the input trajectories $\Sm_1, \dots, \Sm_k$, the optimal $\Gm^*$
defined in~\eqref{eq:opt_kanon} is an elementary partition.
\end{proposition}

\pf Suppose $\Gm \in \Kc$ is not elementary, so that it can be refined to
another partition $\widetilde{\Gm} \in \Kc$. In particular, without loss of
generality, suppose that $\Gm = (\Gc_1,\dots,\Gc_Z)$ and $\widetilde{\Gm} =
\left(\widetilde{\Gc}_1,\dots,\widetilde{\Gc}_{Z+1}\right)$, where
\beq \label{eq:G_vs_Gtilde}
\Gc_i = \left\{
\begin{array}{ll}
\widetilde{\Gc}_i, &  i < Z \\
 \widetilde{\Gc}_Z \cup \widetilde{\Gc}_{Z+1}, & i = Z.
 \end{array} \right.
\eeq

From~\eqref{eq:traj_cost} and~\eqref{eq:G_vs_Gtilde}, the difference between
the costs of $\Gm$ and $\widetilde{\Gm}$ is given by
\beq \label{eq:diff_cost_G_Gtilde}
C(\Gm) - C(\widetilde{\Gm}) = c(\Gc_Z) - c(\widetilde{\Gc}_Z) - c(\widetilde{\Gc}_{Z+1}).
\eeq
%$c(\cdot)$ being defined in~\eqref{eq:sample_cost}.

Since $\Gc_Z$ contains the union of raw samples in $\widetilde{\Gc}_Z$ and
$\widetilde{\Gc}_{Z+1}$, we can apply properties~\eqref{eq:time_cost_prop} and ~\eqref{eq:space_cost_prop} (where~\eqref{eq:time_cost_prop} holds because of time coherence) and obtain:
\begin{eqnarray}
c(\Gc_Z)  &=& c_t(\Gc_Z) c_s(\Gc_Z) \non
%& \geq & \left[\Delta t(\widetilde{\Gc}_Z) + \Delta t(\widetilde{\Gc}_{Z+1}) \right] \cdot \non
%&& \left[ \Delta x(\Gc_Z) + \Delta y(\Gc_Z)\right] \non
& \geq & \left(c_t(\widetilde{\Gc}_Z) + c_t(\widetilde{\Gc}_{Z+1}) \right) c_s(\Gc_Z) \non 
& \geq & c_t(\widetilde{\Gc}_Z)c_s(\widetilde{\Gc}_Z) + c_t(\widetilde{\Gc}_{Z+1})c_s(\widetilde{\Gc}_{Z+1}) \non
& = & c(\widetilde{\Gc}_Z) + c(\widetilde{\Gc}_{Z+1}). \label{eq:ineq_sample_cost}
\end{eqnarray}

Comparing~\eqref{eq:ineq_sample_cost} with~\eqref{eq:diff_cost_G_Gtilde}, we
get that $C(\Gm) \geq C(\widetilde{\Gm})$. Thus, to search for
the optimal $\Gm^*$, we can drop $\Gm$ and keep only $\widetilde{\Gm}$. If
$\widetilde{\Gm}$ is not elementary, then we can find one of its refinements,
and repeat the above steps to drop also $\widetilde{\Gm}$. This way, we can
drop all partitions that are not elementary and be left only with elementary
partitions as $\Gm^*$ candidates.
\qedsymb 

\begin{figure}[tb]
\centering
\includegraphics[width=\columnwidth]{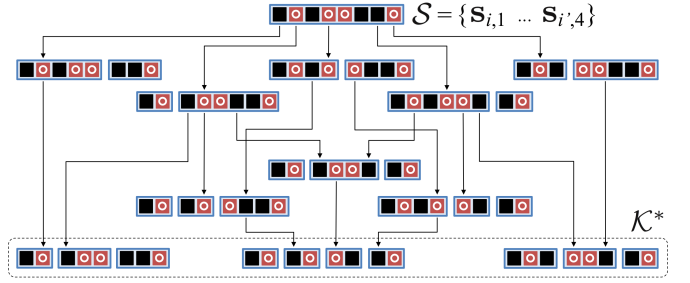}
\caption{Partition tree for the two trajectories
$\Sm_i=\{\sv_{i,j}\}$ and $\Sm_{i'}=\{\sv_{i',j}\}$ in Fig.\,\ref{fig:gen}.
Nodes in the complete tree represent the set $\Kc$ of valid partitions
of the set of raw samples $\Sc$. Elementary partitions are the tree
leaves and constitute $\Kc^*$. The partition in Fig.\,\ref{fig:gen} is
the leftmost leaf in the tree.}
\label{fig:tree}
%\vspace*{-7pt}
\end{figure}

If we build a tree of partitions belonging to $\Kc$, such that the $\Sc$ is the
root and each node is a partition whose children are its refinements, the leaves
are the elementary partitions, which form a subset $\Kc^*$. The above proposition
states that we can limit the search of $\Gm^*$ to $\Kc^*$, drastically reducing
the search space of $\Gm^*$ to the set $\Kc^* \subset \Kc$ of elementary partitions of $\Sc$.
An example is provided in Fig.\,\ref{fig:tree}, for the trajectories in Fig.\,\ref{fig:gen}.

\subsection{Optimal merging algorithm}
%\subsection{\algobase: low-complexity optimal merging of trajectories}
\label{sub:kanon-impl}

We propose \algobase, an algorithm to efficiently search the set of raw samples $\Sc$,
extract the subset of elementary partitions, $\Kc^*$, and identify the optimal
partition $\Gm^*$.
%The algorithm %works for any number $k$ of input trajectories and
%is detailed in Alg.\,\ref{alg:algobase}.

% reduce space after/around algorithm
%\setlength\floatsep{0pt}
%\setlength\textfloatsep{0pt}
%\setlength\intextsep{0pt}

\begin{algorithm}[tb]
\footnotesize
\SetKwInOut{Input}{input}
\SetKwInOut{Output}{output}
\SetKwFunction{timesort}{timesort}
\SetKwFunction{generalize}{generalize}
\SetKwFunction{incomplete}{incomplete}
\SetKwFunction{elementary}{elementary}
\SetKwFunction{visit}{visit}
\SetKwData{Cost}{Cost}
\SetKwData{Partition}{Partition}
\Input{Trajectories $\Sm_1, \dots, \Sm_k$, where $\Sm_i = (\sv_{i,1},\dots,\sv_{i,N_i})$}
\Output{Generalized sample set $\Gm^*$, Cost $C \left(\Gm^*\right)$}
\ForEach{$i \in [1,k]$} { \label{alg1:Sc_start}
	$\Sc_{i} = \bigcup_{j=1}^{N_i} \{\sv_{i,j}\}$\;
}
$\Sc$ $\leftarrow$ \timesort($\Sc_1 \cup \dots \cup \Sc_k$)\; \label{alg1:Sc_end}
%$\Sc$ $\leftarrow$ \timesort($\Sc$)\;
\Cost $\leftarrow$ $(0,\infty,\dots,\infty)$\; %{\scriptsize\tt \% array of |S| elements}
\Partition $\leftarrow$ $(${\sc\scriptsize NULL}, $\dots$, {\sc\scriptsize NULL}$)$\;
\ForEach{$\sv_{\theta} \in \Sc$} { \label{alg1:samples}
	$\theta'=\theta-1$\;
	\While{\incomplete($\sv_{\theta'},\ldots,\sv_\theta$)} { \label{alg1:inc_start}
		$\theta'=\theta'-1$\;
	} \label{alg1:inc_end}
	\While{\elementary($\sv_{\theta'},\ldots,\sv_\theta$)} { \label{alg1:elem_start}
		$\Gc$ $\leftarrow$ \generalize($\sv_{\theta'},\ldots,\sv_\theta$)\; \label{alg1:gen_start}
		\If{\Cost\hspace*{-2pt}$[\theta]$ $> c\left(\Gc\right)$ + \Cost\hspace*{-2pt}$[\theta'-1]$} {
			\Cost\hspace*{-2pt}$[\theta]$ $\leftarrow c\left(\Gc\right)$ + \Cost\hspace*{-2pt}$[\theta'-1]$\; \label{alg1:cost}
			\Partition $\leftarrow (\theta'-1,\Gc)$\; \label{alg1:gen_end2}
		} \label{alg1:gen_end}
		$\theta'=\theta'-1$\;
	} \label{alg1:elem_end}
}
$\Gm^* \leftarrow$ \visit(\Partition)\; \label{alg1:opt_start}
$C\left(\Gm^*\right) \leftarrow $ \Cost\hspace*{-2pt}$[\,|\Sc|\,]$\; \label{alg1:opt_end} % cost of the last element
\caption{\algobase algorithm pseudocode.}
\label{alg:algobase}
\end{algorithm}

The algorithm, detailed in Alg.\,\ref{alg:algobase}, starts by populating
a set of raw samples $\Sc$, whose items
$\sv_{i,j}$ are ordered according to their time value $t(\sv_{i,j})$
(lines~\ref{alg1:Sc_start}--\ref{alg1:Sc_end}).
Then, it processes all samples according to their temporal ordering (line~\ref{alg1:samples}).
Specifically, the algorithm tests, for each sample $\sv_{\theta}$ in position $\theta$, all sets
$\{\sv_{\theta'},\ldots,\sv_{\theta}\}$, with $\theta'< \theta$, as follows.

The first loop skips incomplete sets that do not contain at least one sample from
each input trajectory % fulfil condition \emph{ii)} in Sec.\,\ref{sub:kanon}
(line~\ref{alg1:inc_start}). %(lines~\ref{alg1:inc_start}--\ref{alg1:inc_end}).
The second loop runs until the first non-elementary set is
encountered (line~\ref{alg1:elem_start}). %(lines~\ref{alg1:elem_start}--\ref{alg1:elem_end}).
Therein, the algorithm generalizes the current (complete and elementary)
set $\{\sv_{\theta'},\ldots,\sv_\theta\}$ to $\Gc$, and checks if $\Gc$ reduces the total merging
cost up to $\sv_\theta$. If so, the cost is updated by summing $c(\Gc)$ to the
accumulated cost up to $\sv_{\theta'-1}$, and the resulting (partial) partition of
$\Sc$ that includes $\Gc$ is stored (lines~\ref{alg1:gen_start}--\ref{alg1:gen_end2}).
%This is equivalent to considering $\Sc$ as a graph, where nodes map to
%samples, and edges to partitions in $\Kc^*$.
%
Once out of the loops, the cost associated to the last sample is the optimal cost, and it is
sufficient to backward navigate the partition structure to retrieve the associated $\Gm^*$
(lines~\ref{alg1:opt_start}--\ref{alg1:opt_end}).

Note that, in order to update the cost of including the current sample $\sv_{\theta}$
(line~\ref{alg1:cost}), the algorithm only checks previous samples in time. It thus needs
that the optimal decision up to $\sv_{\theta}$ does not depend on any of the samples in the
original trajectories that come later in time than $\sv_{\theta}$.
The following proposition guarantees that this is the case.
\begin{proposition}
Let $\Gm^* = \left( \Gc^*_1, \dots, \Gc^*_Z \right)$ be the optimal
generalized trajectory and let us make the hypothesis that $\sv_{\theta}$ and $\sv_{\theta+1}$ do not belong to the same generalized sample of $\Gm^*$. Let $\mathbf{G}^*_{\mathrm{p}} =  \left( \Gc^*_1, \dots, \Gc^*_{Z_1} \right)$ and $\mathbf{G}^*_{\mathrm{f}} =  \left( \Gc^*_{Z_1+1}, \dots, \Gc^*_Z \right)$, so that $\sv_{\theta} \in \Gc^*_{Z_1}$ and $\sv_{\theta+1} \in \Gc^*_{Z_1+1}$. Then, $\mathbf{G}^*_{\mathrm{p}}$  can be derived independently of $\mathbf{G}^*_{\mathrm{f}}$.
\end{proposition}
\pf Let $\mathbf{G}$, $\mathbf{G}_{\mathrm{p}}$ and $\mathbf{G}_{\mathrm{f}}$ be  any generalized sequences containing raw samples $(\sv_{1}, \dots, \sv_{N})$, $(\sv_{1}, \dots, \sv_{\theta})$ and  $(\sv_{\theta+1}, \dots, \sv_{N})$, respectively.
According to the cost definition, we generally have 
\begin{eqnarray*}
\min_{\mathbf{G}} C(\mathbf{G}) &\leq&
\min_{\mathbf{G}_{\mathrm{p}},\mathbf{G}_{\mathrm{f}}} C((\mathbf{G}_{\mathrm{p}},\mathbf{G}_{\mathrm{f}}))\\
&=&	\min_{\mathbf{G}_{\mathrm{p}}} C(\mathbf{G}_{\mathrm{p}}) + \min_{\mathbf{G}_{\mathrm{f}}} C(\mathbf{G}_{\mathrm{f}}),
\end{eqnarray*}
where $(\mathbf{G}_{\mathrm{p}},\mathbf{G}_{\mathrm{f}})$ is the concatenation of
$\mathbf{G}_{\mathrm{p}}$ and $\mathbf{G}_{\mathrm{f}}$.
However, by virtue of the hypothesis and by construction, 
\begin{eqnarray*}
\min_{\mathbf{G}} C(\mathbf{G}) &=& C(\Gm^*) \\
&=& C(\mathbf{G}^*_{\mathrm{p}}) + C(\mathbf{G}^*_{\mathrm{f}}) \\
&=& \min_{\mathbf{G}_{\mathrm{p}}} C(\mathbf{G}_{\mathrm{p}}) + \min_{\mathbf{G}_{\mathrm{f}}} C(\mathbf{G}_{\mathrm{f}})
\end{eqnarray*}
so that,  to minimize $C(\mathbf{G})$ we only need to minimize $C(\mathbf{G}_{\mathrm{p}})$ and $C(\mathbf{G}_{\mathrm{f}})$ independently.
\qedsymb

The above proposition guarantees that the algorithm is exploring all
possibilities, and as a result, the cost $C(\Gm^*)$ returned by \algobase is
optimal, i.e., it is the minimum loss of granularity necessary to merge the
original trajectories.%It can thus be also
%intended as an inverse measure of {\it similarity} between spatiotemporal trajectories,
%and we will employ it in this acceptation in the following.

Note that \algobase has a very low complexity in practical cases. Let
$l(\theta)$ be the number of sets $\{\sv_{\theta'},\ldots,\sv_\theta\}$ that are both
complete and elementary for a given $\theta$. Then, the number of computations and
comparisons of sample generalization costs that are performed
in \algobase is $\sum_{\theta} l(\theta) = |\Sc| \overline{l}$, where $\overline{l}$
is the average value of $l(\theta)$. If $\overline{l} = \mathcal O(1)$, which happens
in most trajectory data where the samples of the input trajectories are
%well
intercalated in the time axis, then \algobase runs in a time $\mathcal O(|\Sc|)$,
i.e., linear in the number of samples.

\subsection{Single user \anon}
\label{sub:kte-one}

% --- one-line version
%\begin{figure*}[tb]
%\centering
%\hspace*{-5pt}
%\subfloat[Naive]{
%	\includegraphics[width=0.33\textwidth]{figures/kte-one-naive.eps}\vspace*{3pt}
%	\label{fig:kte-one-naive}
%}
%\subfloat[Overlapping, $\epsilon=\tau$]{
%	\includegraphics[width=0.33\textwidth]{figures/kte-one-2tau.eps}\vspace*{3pt}
%	\label{fig:kte-one-2tau}
%}
%\subfloat[Overlapping, generic]{
%	\includegraphics[width=0.33\textwidth]{figures/kte-one-gen.eps}\vspace*{3pt}
%	\label{fig:kte-one-gen}
%}
%\caption{Implementation of \anon of user $i$.}
%\label{fig:kte-one}
%%\vspace*{-7pt}
%\end{figure*}

% restore space after/around algorithm
%\setlength\floatsep{2\baselineskip}
%\setlength\textfloatsep{2\baselineskip}
%\setlength\intextsep{2\baselineskip}

\begin{figure}[tb]
\centering
% --- EXTENSION
%\vspace*{-5pt}
%\subfloat[Naive]{
%	\includegraphics[width=0.95\columnwidth]{figures/kte-one-naive_trim.eps}\vspace*{3pt}
%	\label{fig:kte-one-naive}
%}\\
%\vspace*{-5pt}
%\subfloat[Overlapping, $\epsilon=\tau$]{
%	\includegraphics[width=0.95\columnwidth]{figures/kte-one-2tau_trim.eps}\vspace*{3pt}
%	\label{fig:kte-one-2tau}
%}\\
%\vspace*{-5pt}
%\subfloat[Overlapping]{
	\includegraphics[width=0.95\columnwidth]{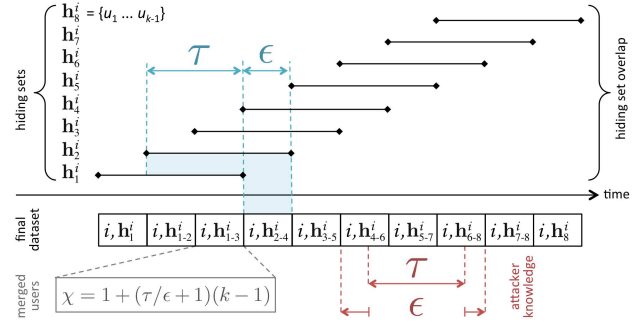}%\vspace*{3pt}
%	\label{fig:kte-one-gen}
%}
\vspace*{-4pt}
\caption{Overlapping hiding set structure realizing \anon for user $i$.}
\label{fig:kte-one-gen}
\vspace*{-10pt}
\end{figure}

%Our objective is in fact attaining not just $k$-anonymity, but the
%more challenging \anon criterion.

%A trivial technique towards \anon would then be to divide time into slots
%of duration $\tau$, and $k$-anonymize $i$ into a different $\hv^i_m$ during
%the $m$-th slot. Unfortunately, this is not a robust solution: we cannot
%assume that the knowledge of the adversary is synchronized with such time
%slots, and a de-synchronized knowledge would reveal the complete
%trajectory of $i$.
%An intuitive example is provided in Fig.\,\ref{fig:kte-one-naive}, where
%the adversary knowledge, indicated in the bottom part of the figure, is in
%between hiding sets $\hv^i_3$ and $\hv^i_4$. The samples known to the attacker
%only partially match the trajectories of users in $\hv^i_3$ and $\hv^i_4$,
%whereas they fully match that of $i$: thus, user $i$ is pinpointed, and his whole
%trajectory becomes known to the adversary.

We implement \anon for a generic subscriber $i$
%by means of overlapping hiding sets
as shown in Fig.\,\ref{fig:kte-one-gen}. 
We discretize time into intervals of length $\epsilon$, named {\it epochs}.
%\footnote{For the sake of clarity, and without loss of generality, we assume
%the adversary knowledge $\tau$ to be a multiple of $\epsilon$.}.
%and, equivalently, $\tau/\epsilon$ to be an integer.}.
At the beginning of the $m$-th epoch, we select a set of $k-1$ users different
from $i$, named a {\it hiding set} of $i$ and denoted as $\hv^i_m$.
The hiding set $\hv^i_m$ provides $k$-anonymity to subscriber $i$ for a
subsequent time window $\tau+\epsilon$.
%: an appropriate generalization makes $i$'s trajectory indistinguishable from those of the users in $\hv^i_m$.
By repeating the hiding set selection for all epochs, $\tau/\epsilon + 1$
subsequent hiding sets of user $i$ overlap at any point in time.
Such a structure of overlapping hiding sets assures the following.

First, subscriber $i$ is $k$-anonymized for any possible knowledge of the
attacker. No matter where a time interval of length $\tau$ is shifted
to along the time dimension, it will be always completely covered by the time
window of one hiding set, i.e., a period during which $i$'s trajectory is
indistinguishable from those of $k-1$ other users.
As an example, in Fig.\,\ref{fig:kte-one-gen}, the attacker knowledge $\tau$
(bottom-right of the plot) is fully enclosed in the time window of $\hv^i_6$,
and his sub-trajectory is indistinguishable from those of users in $\hv^i_6$.

Second, the additional knowledge leaked to the attacker is exactly $\epsilon$.
From the first point above, the adversary cannot tell apart $i$ from the
users in the hiding set $\hv^i_m$ whose time window covers his knowledge
$\tau$. However, the adversary can follow the (generalized) trajectories
of $i$ and users in $\hv^i_m$ for the full time window $\tau+\epsilon$.
Therefore, the adversary can infer new information about the (generalized)
trajectory of $i$ during the time window period that exceeds his original
knowledge $\tau$, i.e., $\epsilon$.
%No matter where the attacker knowledge $\tau$ is shifted to along the time
%dimension, the time window of the hiding set $\hv^i_m$ that covers it exceeds
%$\tau$ by one epoch, i.e., $\epsilon$. 
E.g., in Fig.\,\ref{fig:kte-one-gen}, the time window of $\hv^i_6$
spans before and after the attacker knowledge $\tau$, for a total of $\epsilon$.

The two guarantees above let \anon, as defined in Sec.\,\ref{sub:uninf},
be fulfilled for the generic user $i$. The epoch duration $\epsilon$ maps
to the knowledge leakage.
The following important remarks are in order.

%\vspace*{5pt}%
{\it 1. Hiding set selection.} The structure of overlapping hiding sets is to be
implemented so that the  loss of accuracy in the \anonymized trajectory is minimized.
Thus, the users in the generic hiding set $\hv^i_m$ shall be those who,
during the time window $\tau+\epsilon$ starting at the $m$-th epoch, have
sub-trajectories with minimum \algobase cost with respect to $i$'s.
%To that end, as anticipated in Sec.\,\ref{sub:kanon-impl}, the (inverse
%of the) cost returned by \algobase is an effective similarity measure.
%When considering a single user $i$, one has complete freedom on the choice
%of hiding sets (provided that the reuse constraint is honored, see next
%point), and can pick those users whose sub-trajectories maximize the total
%similarity to $i$'s during each time window $\tau+\epsilon$.

%\vspace*{5pt}%
{\it 2. Reuse constraint.} The uninformative principle requires alternating
the $k-1$ trajectories used in different hiding sets, as per Sec.\,\ref{sub:uninf}.
A simple way to enforce this is limiting the inclusion of any subscriber
in at most one hiding set of $i$.
%%% --- EXTENSION
%This requirement can be relaxed by allowing reuse of the subscribers in multiple hiding sets%
%\footnote{Subscriber reuse is important in practical cases where the number
%of individuals with similar trajectories in the dataset is limited.},
%provided that a {\it reuse constraint} is respected: a same subscriber can be reused across
%hiding sets, but only if such hiding sets are separated by a time at least $\tau$.
%Otherwise, the there is a risk that the attacker can link users in different
%hiding sets, invalidating the leakage bond and thus impairing the
%uninformative principle.
%In conclusion, a subscriber in $\hv^i_m$ can be safely reused to
%$k$-anonymize $i$, but not earlier than in $\hv^i_{m+2\tau/\epsilon}$.

%\vspace*{5pt}%
{\it 3. Generalization set.} As evidenced by the example in Fig.\,\ref{fig:kte-one-gen},
the configuration of hiding sets changes at every epoch, and
$\tau/\epsilon + 1$ hiding sets overlap during each epoch. This means
that a spatiotemporal generalization must be used to merge a
set of $\chi = 1 + (\tau/\epsilon+1)(k-1)$ trajectories at each epoch.

%\vspace*{5pt}%
{\it 4. Epoch duration tradeoff.} The epoch duration $\epsilon$ is a configurable
system parameter, whose setting gives rise to a tradeoff between knowledge leakage
and accuracy of the anonymized data. A lower $\epsilon$ reduces knowledge leakage.
However, it also
%shortens the hiding set time windows $\tau+\epsilon$, or, equivalently,
increases $\chi$,
%A higher $\chi$ means merging $i$'s trajectory with those of a larger number of users,
which typically entails a more marked generalization and a higher loss
of data granularity.

\subsection{Multiple user \anon}
\label{sub:kte-mul}

Scaling \anon from a single user to all subscribers in a dataset implies that
the choice of hiding sets cannot be made independently for every user.
Therefore, trajectory similarity and reuse constraint fulfillment are not
sufficient norms anymore.
In addition to the above, the selection of hiding sets needs to
be concerted among all users so as to ensure that the generalized trajectories
are correctly intertwined and all subscribers are $k$-anonymized during each
time window $\tau+\epsilon$.

% --- NOTE: full consitency is impossible at epoch level (i.e., trying
%     to have ABC + BAC + CAB at each epoch, for 2^{\tau,2}-anonymity),
%     but it is feasible on a hiding set basis (i.e., AB + BA for a
%     2^{\tau,2}-anonymity). This second case is in fact one way to
%     satisfy the one-pick constraint on each epoch!
An intuitive solution is enforcing {\it full consistency}: including a
subscriber $i$ into the hiding set of user $i'$ at epoch $m$ makes $i'$
automatically become part of $i$'s hiding set at the same epoch.
Formally, $i\in\hv^{i'}_m \Rightarrow i'\in\hv^{i}_m$, $\forall i \neq i', \forall m$.

\begin{figure}[tb]
\centering
\includegraphics[width=0.9\columnwidth]{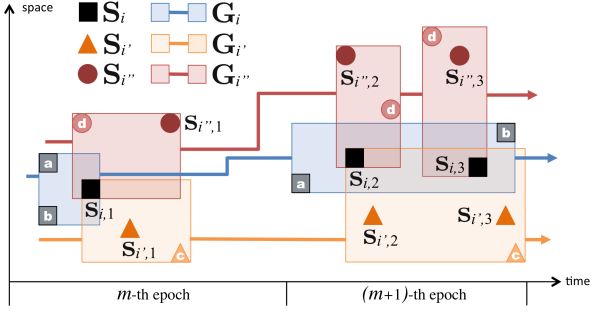}
\vspace*{-4pt}
\caption{Example of $k$-pick constraint, with $k$=3, for user $i$
during the $m$-th hiding set selection. Here $\epsilon = \tau$, hence
the time windows of hiding sets span two epochs.
For clarity, space is unidimensional.
Figure best viewed in colors.}
\label{fig:kte-multi}
\vspace*{-10pt}
\end{figure}

In fact, full consistency is an unnecessarily restrictive condition. It is
sufficient that hiding set concertation satisfies a {\it $k$-pick constraint}:
during the $m$-th epoch, each user $i$ in the dataset has to be picked in the
hiding sets of at least other $k-1$ subscribers. Formally, $|\{i', \;
i\in\hv^{i'}_m\}| \geq k-1$, $\forall i, \forall m$. This provides an
increased flexibility over all existing approaches which rely on fully
consistent generalization strategies.

The rationale behind the $k$-pick constraint is best illustrated by means
of a toy example, in Fig.\,\ref{fig:kte-multi}.
The figure portrays the spatiotemporal samples of users $i$, $i'$ and $i''$
during epochs $m$ and $m+1$. The sub-trajectory of subscriber $i$ in this
time interval is $\Sm_i = (\sv_{i,1},\sv_{i,2},\sv_{i,3})$, represented as
black squares; equivalently for $i'$ (orange triangles) and $i''$ (red circles).
Samples denoted by letters belong to other users $a$, $b$, $c$ and $d$, and 
they are instrumental to our example. %, but do not need to be defined further.

Let us assume that $\epsilon=\tau$ (i.e., hiding sets span an interval $2\tau=2\epsilon$,
or epochs $m$ and $m+1$), and $k=3$. At the beginning of the $m$-th epoch, for
subscriber $i$ (resp., $i'$ and $i''$), one needs to select $k-1=2$ other
users that constitute the hiding set $\hv^i_m$ (resp., $\hv^{i'}_m$
and $\hv^{i''}_m$).
Let us consider $\hv^i_m$=$\{a,b\}$, $\hv^{i'}_m$=$\{i,c\}$, $\hv^{i''}_m$=$\{i,d\}$,
which results in the generalized sub-trajectories $\Gc_i$, $\Gc_{i'}$, $\Gc_{i''}$
in Fig.\,\ref{fig:kte-multi}.
%%% --- EXTENSION
%It easy to see that this selection minimizes the merging cost, i.e., the loss
%of granularity.
%
%More importantly,
The configuration satisfies the $k$-pick constraint for
subscriber $i$, who is picked in $k-1=2$ hiding sets, i.e., $\hv^{i'}_m$
and $\hv^{i''}_m$.
Suppose now that the attacker knows the spatiotemporal samples of $i$'s
trajectory during any time interval $\tau$ within the $m$-th and $(m+1)$-th
epoch: as these samples are within $\Gc_i$, $\Gc_{i'}$ and $\Gc_{i''}$,
then $i$ is
%successfully
$3$-anonymized.

%Full consistency would now require that $x$ (respectively, $y$) pick $i$
%and $y$ (respectively, $x$) in their hiding sets, i.e., that
%$\hv^i_m = \hv^x_m = \hv^y_m = \left\{i,x,y\right\}$.

The key consideration is that $i$ is $k$-anonymized at epoch $m$ by $i'$ and
$i''$, yet it does not contribute to the anonymization of neither $i'$ nor
$i''$, as $i',i''\notin\hv^i_m$. Thus, it is possible to decouple the choice
of hiding sets across subscribers, without jeopardizing the privacy guarantees
granted by $k$-anonymity.
Such a decoupling entails a dramatic increase of flexibility in the choice of
hiding sets, as per the following proposition.
\begin{proposition}
Given a dataset of $U$ trajectories and a fixed value of $k$, the number of hiding set configurations allowed by
full consistency is a fraction of that allowed by $k$-pick that vanishes more than exponentially for $U \to \infty$.
\end{proposition}
\pf 
Let us consider a set of $U$ users, where $U$ is a multiple of $k$, since otherwise
full consistency cannot even be enforced. Let us build a $k \times U$ matrix, in which the $i$-th column contains $(i, \hv^i_m)$, where $\hv^i_m$ is the  hiding set for user $i$ at a given epoch $m$. (For simplicity, in this proof, we do not take into account the reuse constraints.)

The solution set under the
$k$-pick constraint coincides with the set of normalized Latin rectangles\footnote{A $k \times n$ \emph{Latin rectangle}, $k \leq n$, is a matrix in which all entries are taken from the set $\{1,\dots,n\}$, in such a way that each row and column contains each value at most once. The Latin rectangle is said to be normalized if the first row is the ordered set $(1,\dots,n)$.} of size $k \times U$. Let $K_{k,U}$ be the number of $k \times U$ normalized Latin rectangles, which equals the number of possible solutions for our problem with the $k$-pick constraint. An old result by Erd\H{o}s and Kaplansky~\cite{erdos} states that, for $U \to \infty$ and $k = O \left((\log U)^{3/2-\epsilon})\right)$, 
\beq
K_{k,U} \sim (U!)^{k-1} \exp \left( -k(k-1)/2 \right)
\eeq

If, instead, we enforce full consistency, then the number of solutions equals the number of different partitions of a size-$U$ set into $U/k$ subsets, all with size $k$. Denoting by $C_{k,U}$ this number, we can compute it as
\beq
C_{k,U} = \frac{{U \choose k}{U-k \choose k}\cdot \dots \cdot {k \choose k}}{\left(U/k\right)!} = \frac{U!}{(k!)^{U/k} \left(U/k\right)!}
\eeq

Thus, for fixed $k$ and $U \to \infty$
\[
\frac{C_{k,U}}{K_{k,U}} \sim \frac{\exp \left( k(k-1)/2 \right)}{(U!)^{k-2}(k!)^{U/k} \left(U/k\right)!}
\]
which tends to zero more than exponentially for $U \to \infty$.
\qedsymb

For large datasets of hundreds of thousands trajectories, $k$-pick enables
a much richer choice of merging configurations. This reasonably unbinds
better combinations of the original trajectories, and results in more
accurate anonymized data.

\subsection{Practical \anon algorithm}
\label{sub:algo}

Capitalizing on all previous results, we design \algokte, an algorithm
that achieves \anon in datasets of spatiotemporal trajectories.
Since even the optimal solution to the simpler $k$-anonymity problem is
known to be NP-hard~\cite{fung10}, we resort here to an heuristic solution.

% reduce space after/around algorithm
%\setlength\floatsep{0pt} %.25\baselineskip}
%\setlength\textfloatsep{0pt} %.25\baselineskip}
%\setlength\intextsep{0pt} %.25\baselineskip}

\begin{algorithm}[tb]
\footnotesize
\SetArgSty{}
\SetKwInOut{Input}{input}
\SetKwInOut{Output}{output}
\SetKwFunction{epochs}{epochs}
\SetKwFunction{filter}{filter}
\SetKwFunction{spectralClustering}{spectralClustering}
\SetKwFunction{split}{split}
\SetKwFunction{graph}{graph}
\SetKwFunction{greedyCycle}{greedyCycle}
\SetKwFunction{suppression}{suppression}
\SetKwFunction{merge}{merge}
\SetKwFunction{replace}{replace}
\SetKwData{Costs}{Costs}
\SetKwData{Clusters}{Clusters}
\SetKwData{Subclusters}{Subs}
\Input{Anonymization level $k$, attacker knowledge $\tau$, leakage $\epsilon$}
\Input{Trajectory dataset $\mathbb{D}$}
\Output{Anonymized trajectory dataset $\mathbb{D}$}
\ForEach{$e_\theta \in$ \epochs($\mathbb{D}$)} { \label{alg2:epoch}
	$\mathbb{D}_f$ $\leftarrow$ \filter($e_\theta,\mathbb{D}$)\; \label{alg2:epoch_start}
	\ForEach{$\Sm_i,\Sm_{i'} \in \mathbb{D}_f, \Sm_i \neq \Sm_{i'}$} {
		\Costs[$\Sm_i,\Sm_{i'}$] $\leftarrow$ \algobase($\Sm_i,\Sm_{i'}$)\; \label{alg2:epoch_end2}
	} \label{alg2:epoch_end}
	\Clusters[$\theta$] $\leftarrow$ \spectralClustering($\Costs$)\; \label{alg2:sc}
  \If{$\theta \ge \tau/\epsilon+1$} { \label{alg2:epoch_min}
		\ForEach{$\texttt{c} \in$ \Clusters[$\theta$]} { \label{alg2:cluster}
			\Subclusters $\leftarrow$ \split($\texttt{c}$,\Clusters[$\theta-\tau/\epsilon$ : $\theta-1$])\; \label{alg2:split}
			\ForEach{$\texttt{c}_s \in$ \Subclusters[$\theta$]} {
				$\texttt{g}_{s} \leftarrow$ \graph($\texttt{c}_s$)\; \label{alg2:graph}
				$\texttt{g}_{sc} \leftarrow$ \greedyCycle($\texttt{g}_{s}$,$k$)\; \label{alg2:greedycycle}
				\If{$\exists \texttt{g}_{sc}$} { 
					\ForEach{$\Sm_i \in \texttt{c}_s$} { \label{alg2:hv_start}
						$\hv^i_{\theta-\tau/\epsilon} \leftarrow \texttt{g}_{sc}$[$\Sm_i$]\; \label{alg2:hv_end2}
					}\label{alg2:hv_end}
				} 
				\Else {
					\suppression($\texttt{c}_s$)\; \label{alg2:suppression}
				}
			}
		}
	}
}
\ForEach{$e_\theta \in$ \epochs($\mathbb{D}$)} { \label{alg2:merge_start}
	\ForEach{$\Sm_i \in \mathbb{D}$} {
		$\hv \leftarrow$ \filter($e_\theta,\Sm_i,\hv^i_{\theta-\tau/\epsilon},\dots,\hv^i_\theta$)\;
		$\mathbb{D} \leftarrow$ \replace(\algobase($\hv$))\; \label{alg2:merge_end2}
	}
} \label{alg2:merge_end}
\caption{\algokte algorithm pseudocode.}
\label{alg:algokte}
\end{algorithm}

The algorithm, in Alg.\,\ref{alg:algokte}, proceeds on a per-epoch basis (line~\ref{alg2:epoch}),
finding, for each epoch $\theta$, a set of $\chi$ users (with $\chi$
defined as in Sec.\,\ref{sub:kte-one}) that hide each subscriber at
low merging cost.
An extensive search for the set of $\chi$ users would have an excessive
cost $\Oc(U^\chi)$, where $U$ is the number of users in dataset, and
$\chi\geq3$. Thus, we adopt a computationally efficient approach, by
clustering user sub-trajectories based on their pairwise merging cost.
Costs are computed via \algobase
(lines~\ref{alg2:epoch_start}--\ref{alg2:epoch_end2}), and a standard
spectral clustering algorithm
%~\cite{ng02}
groups similar trajectories
into same clusters (line~\ref{alg2:sc}). This allows operating on each
cluster independently in the following.

Starting from epoch $\tau/\epsilon+1$ (line~\ref{alg2:epoch_min}),
the algorithm processes each identified cluster at epoch $\theta$
separately (line~\ref{alg2:cluster}).
It splits the current cluster $c$ into subsets, which contain
user trajectories that share the same sequence of clusters during
the last $\tau/\epsilon$ epochs (line~\ref{alg2:split}).

Let $c_s$ be any of such subsets:
%All users within $c_s$ have similar
%sub-trajectories during the previous interval of length $\tau$.
$c_s$ is mapped to a directed graph whose nodes are the
users within $c_s$, and there is an edge going from user $j$ to user
$i$ if $j$ can be in the hiding set $\hv^i_{\theta-\tau/\epsilon}$
of $i$  without violating the reuse constraint (line~\ref{alg2:graph}).
If a $k$-anonymity level is required, $k-1$ directional cycles are then
built within the graph, involving all nodes in the graph, in such a way
that each node has a different parent in each cycle (line~\ref{alg2:greedycycle}).
The hiding set $\hv^i_{t-\tau/\epsilon}$ is then obtained as
the set of user $i$'s parents in the $k-1$ cycles
(lines~\ref{alg2:hv_start}--\ref{alg2:hv_end2}).

Such a construction  of hiding sets complies with the $k$-pick constraint,
since every user $i$ is in the hiding set of $k-1$ other users. 
It may however happen that no valid $k-1$ cycles can be created within $c_s$:
this means that subscribers in $c_s$ share a sub-trajectory
that is rare in the dataset, and their number is insufficient to implement
\anon. In this case, we apply suppression and remove all spatiotemporal
samples of such users' sub-trajectories (line~\ref{alg2:suppression}).
Once all hiding sets are determined, the merging is performed, on each epoch
and for each user, using \algobase (lines~\ref{alg2:merge_start}--\ref{alg2:merge_end2}).

Overall, the heuristic algorithm above guarantees that overlapping hiding sets
that satisfy the reuse constraint (Sec.\,\ref{sub:kte-one}) are selected for all
users. It also ensures that such a choice of hiding sets fulfils the $k$-pick
requirement (Sec.\,\ref{sub:kte-mul}). Together, these conditions realize \anon
of the trajectory data.

%We calculate the complexity of \algokte by analyzing its subroutines, 
%recalling the complexity of \algobase detailed in Sec.~\ref{sub:kanon-impl}.
The complexity of \algokte is as follows.
Let $U$ be the number of users, $\Theta$ be the number of epochs and $\overline{\Nc}$
be the average number of samples per user per epoch, so that
$\Nc_{tot} = \Theta U \overline{\Nc}$ is the total number of samples in the
dataset. Then:
%\vspace*{1pt}
{\it (i)}
lines~\ref{alg2:epoch_start}--\ref{alg2:epoch_end2}
perform \algobase on two input trajectories $\Theta U^2 $ times, each of them
with a complexity $\Oc(\overline{\Nc} )$, for a total complexity of 
$\Oc(\Nc_{tot} U )$;
%\vspace*{1pt}
{\it (ii)}
%
%Our implementation of spectral clustering (line~\ref{alg2:sc}), which
%uses the implicitly restarted Lanczos method~\cite{saad11} to find the $m$
%largest eigenvalues of the similarity matrix, with $m = \sqrt{U}$, has a
%complexity of $\Oc(\Theta U^2\sqrt{U} )$%
%\footnote{There exist suboptimal implementations of spectral clustering
%(like KASP~\cite{yan09}) that allow reducing the complexity to $\Oc(\Theta U^2 )$.};
spectral clustering (line~\ref{alg2:sc}) can be implemented with complexity
$\Oc(\Theta U^2 )$ using KASP~\cite{yan09};
%\vspace*{1pt}
{\it (iii)}
the complexity of lines~\ref{alg2:merge_start}--\ref{alg2:merge_end2}, performing
\algobase on $\chi$ input trajectories $\Theta U $ times, is $\Oc(\Nc_{tot} \chi )$.
%\vspace*{1pt}
All other subroutines of \algokte have a much smaller complexity.

\begin{table*}[tb]
\centering
\fontsize{6}{2}\selectfont%
\renewcommand{\arraystretch}{4}
\setlength{\tabcolsep}{2.5pt}
\begin{minipage}[t]{0.41\textwidth}
\centering
\caption{Features of reference mobile traffic datasets.}
\vspace*{-7pt}
\label{tbl:dataset_stats}
\begin{tabular}{|c|r|r|r|r|r|r|r|}
\hline
Dataset & Surface & BS & BS/Km\textsuperscript{2} & Users & Density       &  Samples            & Timespan \\
        & {[}Km\textsuperscript{2}{]} &    &           &       & {[}user/Km\textsuperscript{2}{]} & {[}per user/h{]} & {[}days{]} \\
\hline
\texttt{abi} & 2,731 & 400  & 0.14 & 29,191 & 10.68 & 0.90 & 14 \\
\hline
\texttt{dak} & 1,024 & 457  & 0.44 & 71,146 & 69,47 & 0.74 & 14 \\
\hline
\texttt{shn} & 3,329 & 2961 & 0.89 & 50,000 & 15.01 & 1.00 & 1 \\
\hline
\texttt{civ} & 322,463 & 1238  & 0.0038 & 82,728 & 0.26 & 0.75 & 14 \\
\hline
\texttt{sen} & 196,712 & 1666 & 0.0085 & 286,926 & 1.45 & 0.45 & 14 \\
\hline
\end{tabular}
\end{minipage}
\hfill
\begin{minipage}[t]{0.57\textwidth}
\centering
\caption{Comparative performance evaluation of \algobase} %for spatiotemporal trajectory merging.}
\vspace*{-7pt}
\label{tab:algobase}
\begin{tabular}{|c|c|r|r|r|r|r|r|r|r|r|r|r|}
\hline
\multirow{3}{*}{Dataset}              & \multirow{3}{*}{$k$} & \multicolumn{2}{c|}{\algobase} & \multicolumn{3}{c|}{Static generalization {[}success \%{]}} & \multicolumn{4}{c|}{W4M}                             & \multicolumn{2}{c|}{GLOVE} \\
\cline{3-13}
                                      &   & Time & Space & 2h - 4Km         & 4h - 10Km        & 8h - 20Km  & Deleted & Created & Time & Space &  Time & Space     \\
%\cline{3-13}
                                      &   & {[}min{]} & {[}Km{]} &          &                  &            & {[}\%{]} & {[}\%{]} & {[}min{]} & {[}Km{]} &  {[}min{]} & {[}Km{]}     \\
\hline
\multirow{3}{*}{\texttt{abi}}     & 2                  & 51            & 0.624           & 27.2              & 56.7             & 80.3        & 9.6              & 22.0                & 57            & 1.166          & 114 &  2.626 \\ \cline{2-13} 
											   & 5                  & 228           & 3.423          & 0.7               & 11.0               & 40.5    & 31.9             & 31.2              & 185           & 3.809          & 292 & 3.740          \\ \cline{2-13} 
                         & 8                  & 349           & 5.720          & 0.1               & 5.1              & 22.6       & 23.9             & 36.7              & 198           & 6.163          & --- &  ---    \\ \hline
\multirow{3}{*}{\texttt{dak}}     & 2                  & 47            & 0.701          & 43.2              & 68.7             & 93.3     & 5.9              & 11.4              & 39            & 1.466          & 116 & 2.498    \\ \cline{2-13} 
                         & 5                  & 220           & 5.286          & 2.2               & 14.0             & 67.0         & 20.3             & 21.2              & 172           & 5.807          & 294 & 3.192   \\ \cline{2-13} 
                         & 8                  & 377           & 7.794          & 0.1               & 8.6              & 50.7        & 22.0               & 18.6              & 189           & 8.477          & --- &  ---   \\ \hline
\end{tabular}
\end{minipage}
\end{table*}

\section{Performance evaluation}
\label{sec:results}

We evaluate our anonymization solutions with five real-world
datasets of mobile subscriber trajectories, introduced in
Sec.\,\ref{sub:datasets}. A comparative evaluation of \algobase
is in Sec.\,\ref{sub:peva-algobase}, while the results of
\anonymization via \algokte are presented in Sec.\,\ref{sub:peva-algokte}. 

\subsection{Reference datasets}
\label{sub:datasets}

%%% --- EXTENSION
% The extended version of the results is in the submitted version of Sigcomm'16.

Our datasets consist of user trajectories extracted from call detail records
(CDR) released by Orange within their D4D Challenges~\cite{d4d}, and by the
University of Minnesota~\cite{zhang14}.
Three datasets, denoted as \texttt{abi}, \texttt{dak} and \texttt{shn},
describe the spatiotemporal trajectories of tens of thousands
mobile subscribers in urban regions, while the other two, \texttt{civ}
and \texttt{sen} hereinafter, are nationwide.
In all datasets, user positions map to the latitude and longitude
of the current base station (BS) they are associated to.
The main features of the datasets are listed in Tab.\,\ref{tbl:dataset_stats},
revealing the heterogeneity of the scenarios.

In order to ensure that all datasets yield a minimum level of
detail in the trajectory of each tracked subscriber, we had to
preprocess the \texttt{abi} and \texttt{civ} datasets. Specifically,
we only retained those users whose trajectories have at least one
spatiotemporal sample on every day in a specific two-week period.
No filtering was needed for the \texttt{dak} and \texttt{sen} datasets,
which already contain users who are active for more than 75\% of a
2-week timespan, and \texttt{shn}, whose users have even higher
sampling rates.

In all datasets, user positions map to the latitude and longitude of the
current base station (BS) they are associated to. We discretized the resulting
positions on a 100-m regular grid, which represents the finest spatial granularity
we consider%
\footnote{At 100-m spatial granularity, each grid cell contains at most one antenna
from the original dataset: the process does not cause any loss in data accuracy.}.

Samples are timestamped with an precision of one minute. This is the granularity
granted in the \texttt{abi} and \texttt{civ} datasets. The \texttt{dak} and \texttt{sen}
datasets feature a temporal granularity of 10 minutes: in order to have comparable
datasets, we added a random uniform noise over a ten-minute timespan to each sample,
so as to artificially refine the time granularity of the data to one minute as well.
In the case of the \texttt{shn} dataset, the precision is one second, and we used
a one-minute binning to uniform the data to the standard format.

\subsection{Comparative evaluation of \algobase}
\label{sub:peva-algobase}

Since no previous solution for \anon exists, we are forced to compare our
algorithms to previous techniques in terms of simpler $k$-anonymity.
Interestingly, this allows validating our proposed approach for merging
spatiotemporal trajectories via the \algobase algorithm.

We thus run \algobase on 100 random $k$-tuples of mobile users from the
reference datasets, for different values of $k$, and we record the
spatiotemporal granularity retained by the resulting generalized trajectories.
We compare our results against those obtained by the only three approaches
proposed in the literature for the $k$-anonymization of trajectories along both
spatial and temporal dimensions.

The first is static generalization~\cite{zang11,de-montjoye13}, which consists in
a homogeneous reduction of data granularity, decided arbitrarily and imposed on
all user trajectories. Static generalization is a trial-and-error process, and
it does not guarantee $k$-anonymity of all users.
The second benchmark solution is Wait for Me (W4M)~\cite{abul10}. Intended for
regularly sampled (e.g., GPS) trajectories, W4M performs the minimum spatiotemporal
translation needed to push all the trajectories  within the same cylindrical volume.
It allows the creation of new synthetic samples, and it is thus not fully compliant
with PPDP principles in Sec.\,\ref{sub:ppdp}.
The latter operation is leveraged to improve the matching among trajectories in a
cluster, and assumes that mobile objects (i.e., subscribers in our case) effectuate
linear constant-speed movements between spatiotemporal samples. We use W4M with
linear spatiotemporal distance (W4M-L), i.e., the version intended for large databases
such as those we consider
\footnote{Implementation at {\tt http://kdd.isti.cnr.it/W4M/}.},
and configure it with the settings suggested in~\cite{abul10}.
The third approach is GLOVE~\cite{gramaglia15}, which relies on a heuristic
measure of anonymizability to assess the similarity of spatiotemporal trajectories.
This measure is fed to a greedy algorithm to achieve $k$-anonymity with limited loss
of granularity and without introducing fictitious data. However, unlike \algobase,
GLOVE does not provide an optimal solution, and is computationally expensive.

The results of our comparative evaluation are summarized in Tab.\,\ref{tab:algobase},
for the \texttt{abi} and \texttt{dak} datasets, when varying number $k$ of
trajectories merged together. Similar results were obtained for the other
datasets, and are omitted due to space limitations.
We immediately note how static aggregation is an ineffective approach: the percentage
of successfully merged $k$-tuples is well below 100\%, even when dramatically reducing
the data granularity to 8 hours in time and 20 km in space.
Instead, \algobase, W4M and GLOVE can merge all of the $k$-tuples, while
retaining a good level of accuracy in the data.
We can directly compare the granularity in time (min) and space (km) retained
by \algobase, W4M and GLOVE in merging groups of $k$ trajectories: the
spatiotemporal accuracy is comparable in all cases.
However, it is important to note that W4M attains this result by deleting and creating
a significant amount of samples: in the end, only 40-70\% of the original samples are
maintained in the generalized data. Conversely, all of the generalized samples
created by \algobase reflect the actual real-world data.
Also, \algobase obtains a level of precision that is always higher than that
of GLOVE, and scales better: indeed, the complexity of GLOVE did not allow
computing a solution when $k=8$.

Overall, the results uphold \algobase as the current state-of-the-art solution
to generalize sparse spatiotemporal trajectories while obeying PPDP principles
and minimizing accuracy loss.
%in the data.

% --- EXTENSION
%\begin{figure}[tb]
%\centering
%\subfloat[abi]{\label{fig:pos_gran_cdf_abi}
%   \includegraphics[width=0.229\textwidth]{figures/d4d13_abi_pos_gran_cdf.eps}
%}
%\subfloat[dak]{\label{fig:pos_gran_cdf_dak}
%   \includegraphics[width=0.229\textwidth]{figures/d4d14_dak_pos_gran_cdf.eps}
%}
%\caption{pos gran cdf}
%\label{fig:pos_gran_cdf}
%\end{figure}
%
%\begin{figure}[tb]
%\centering
%\subfloat[abi]{\label{fig:time_gran_cdf_abi}
%   \includegraphics[width=0.229\textwidth]{figures/d4d13_abi_time_gran_cdf.eps}
%}
%\subfloat[dak]{\label{fig:time_gran_cdf_dak}
%   \includegraphics[width=0.229\textwidth]{figures/d4d14_dak_time_gran_cdf.eps}
%}
%\caption{time gran cdf}
%\label{fig:time_gran_cdf}
%\end{figure}

%\begin{figure}[tb]
%centering
%\subfloat[\texttt{d4d-abi}]{\label{fig:pos_gran_vs_t_abi}
%		\includegraphics[width=0.23\textwidth]{figures/d4d13_abi_pos_gran_vs_t.eps}
%}
%\subfloat[\texttt{d4d-dak}]{\label{fig:pos_gran_vs_t_dak}
%   \includegraphics[width=0.23\textwidth]{figures/d4d14_dak_pos_gran_vs_t.eps}
%}
%\caption{Spatial granularity versus adversary knowledge.}
%\label{fig:pos_gran_vs_t}
%\end{figure}

%\begin{figure}[tb]
%\centering
%\subfloat[\texttt{d4d-abi}]{\label{fig:tim_gran_vs_t_abi}
%   \includegraphics[width=0.23\textwidth]{figures/d4d13_abi_tim_gran_vs_t.eps}
%}
%\subfloat[\texttt{d4d-dak}]{\label{fig:tim_gran_vs_t_dak}
%   \includegraphics[width=0.23\textwidth]{figures/d4d14_dak_tim_gran_vs_t.eps}
%}
%\caption{Temporal granularity versus adversary knowledge.}
%\label{fig:tim_gran_vs_t}
%\end{figure}

\begin{figure*}[tb]
\captionsetup[subfloat]{captionskip=-3pt}
\centering
\hspace*{-8pt}
\subfloat[\texttt{abi}]{\label{fig:pos_gran_vs_t_abi}
   \includegraphics[width=0.22\textwidth]{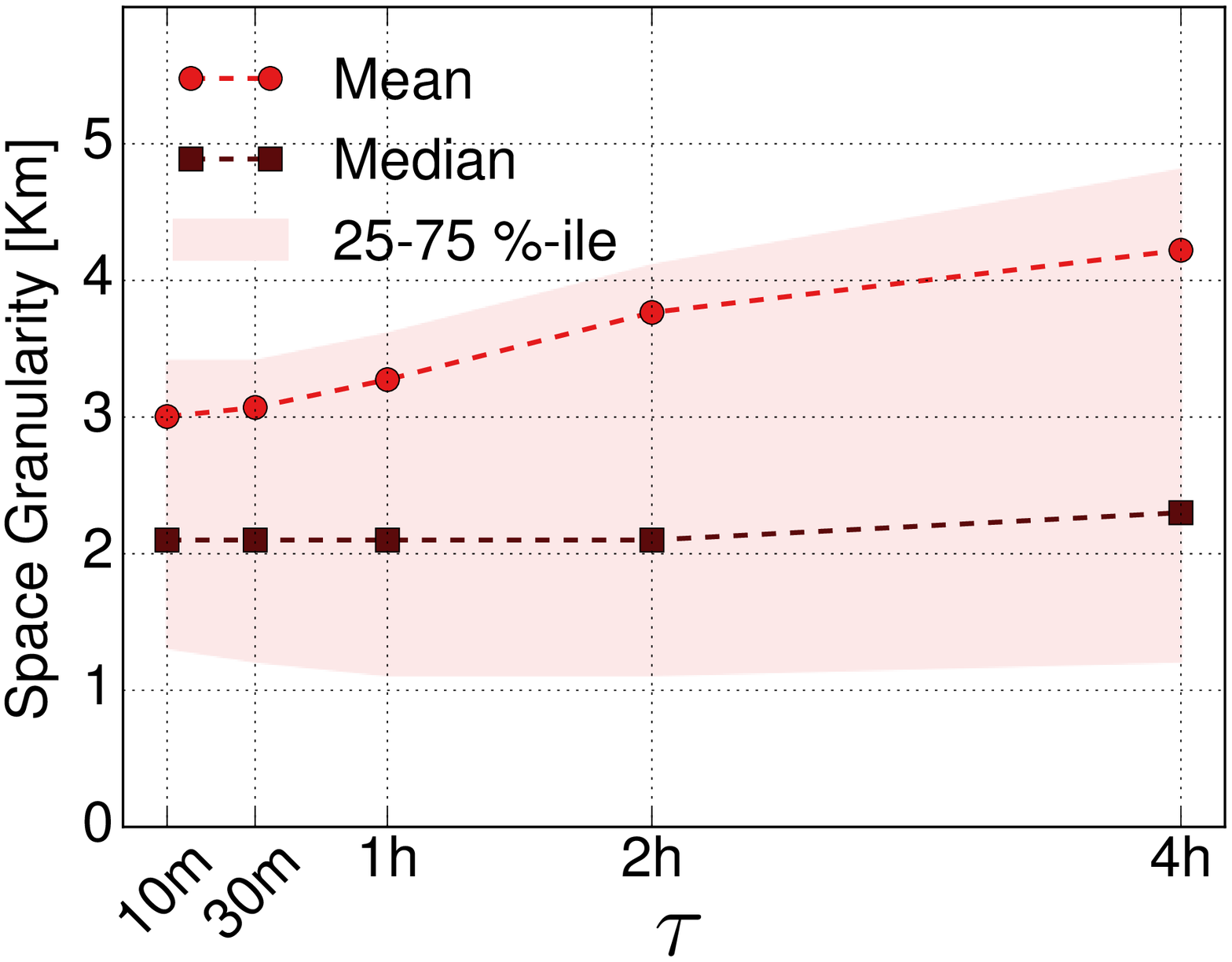}
}
\hspace*{-8pt}
\subfloat[\texttt{dak}]{\label{fig:pos_gran_vs_t_dak}
   \includegraphics[width=0.218\textwidth]{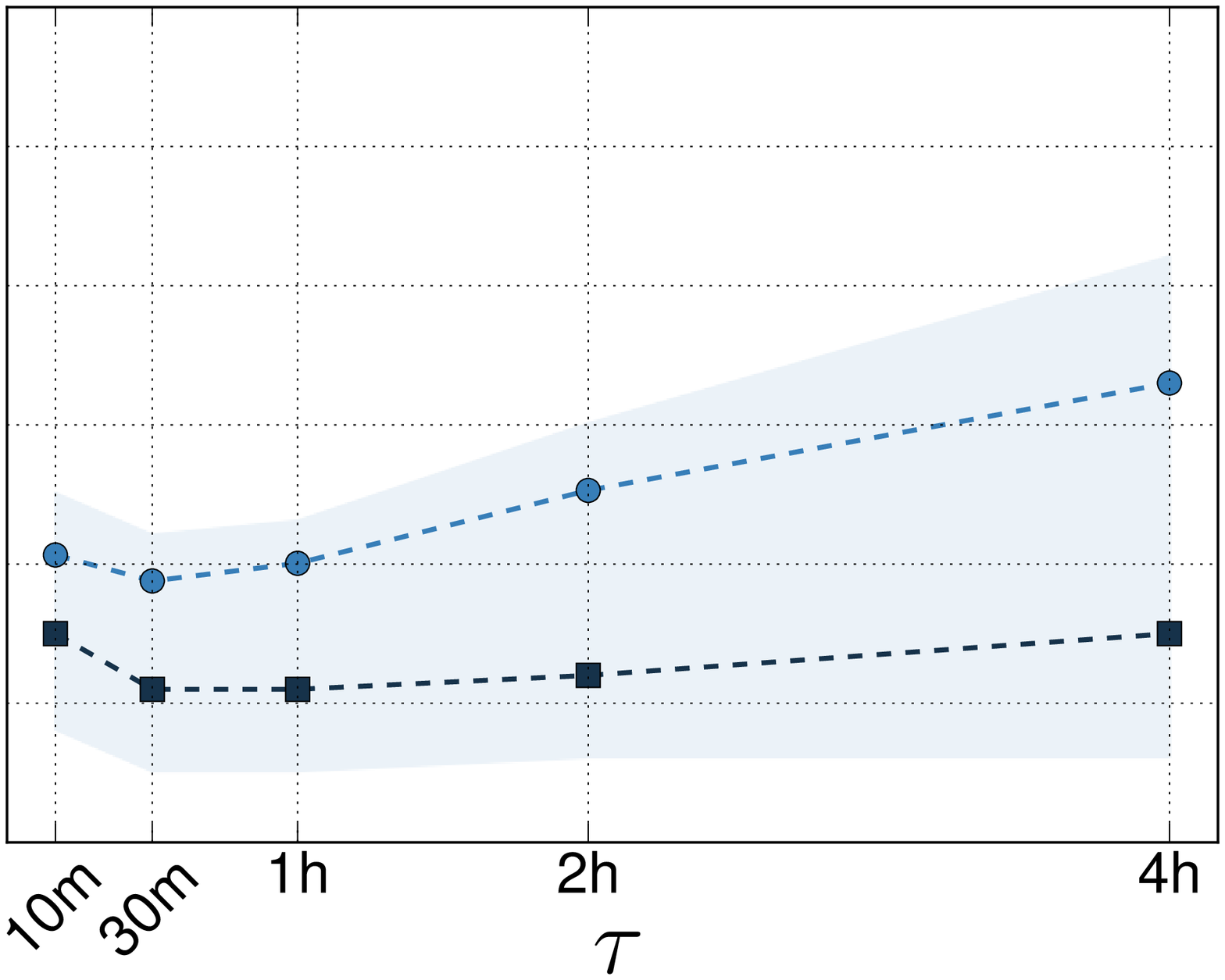}
}
\hspace*{-10.7pt}
\subfloat[\texttt{shn}]{\label{fig:pos_gran_vs_t_shn}
   \includegraphics[width=0.0624\textwidth]{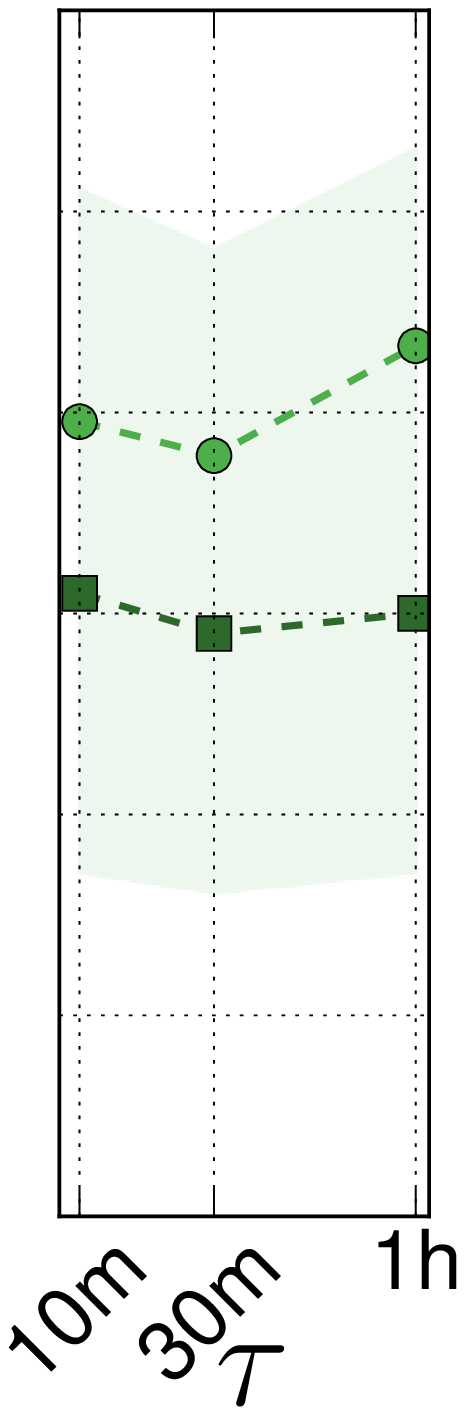}
}
\hspace*{-8pt}
\subfloat[\texttt{abi}]{\label{fig:tim_gran_vs_t_abi}
   \includegraphics[width=0.22\textwidth]{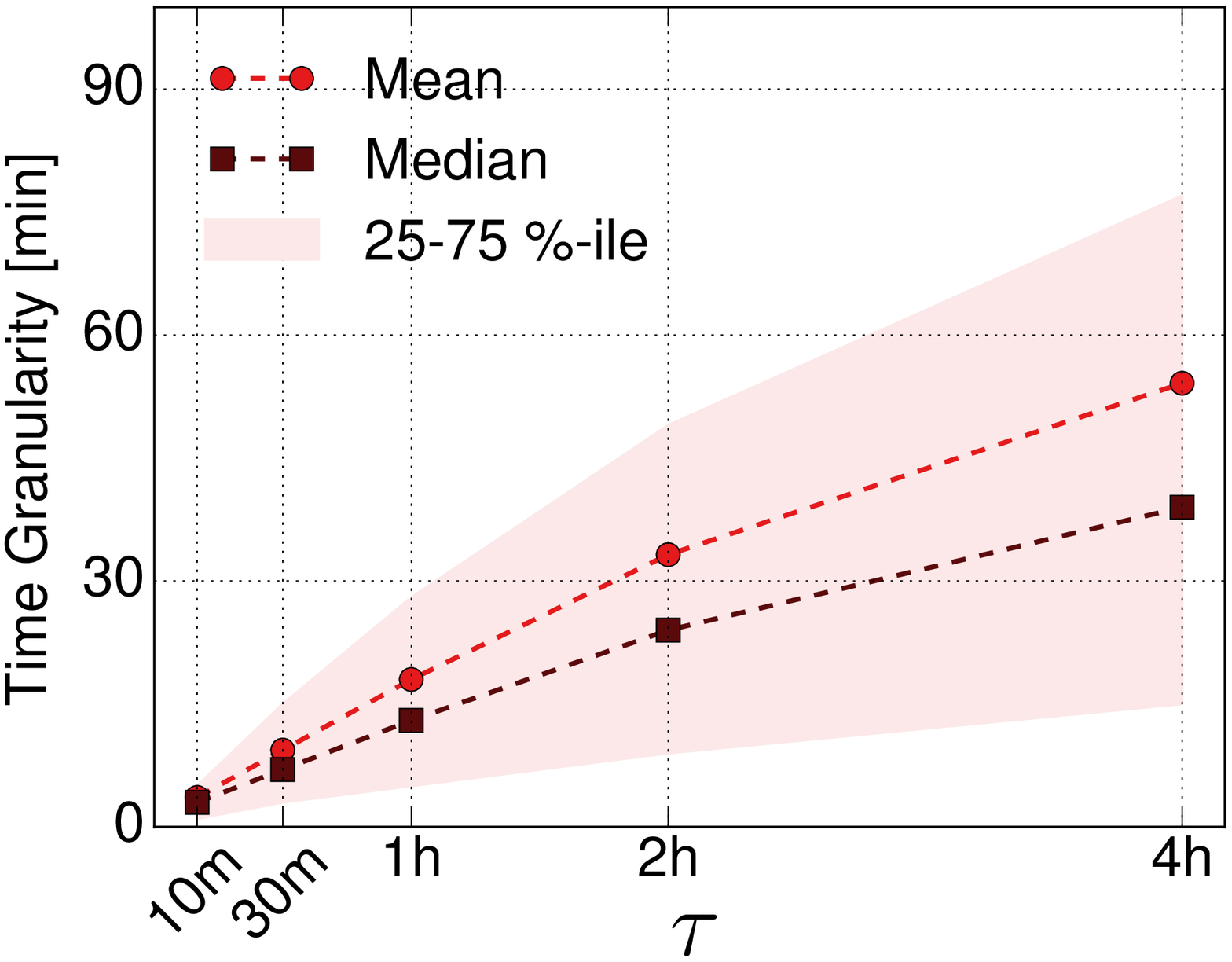}
}
\hspace*{-8pt}
\subfloat[\texttt{dak}]{\label{fig:tim_gran_vs_t_dak}
   \includegraphics[width=0.218\textwidth]{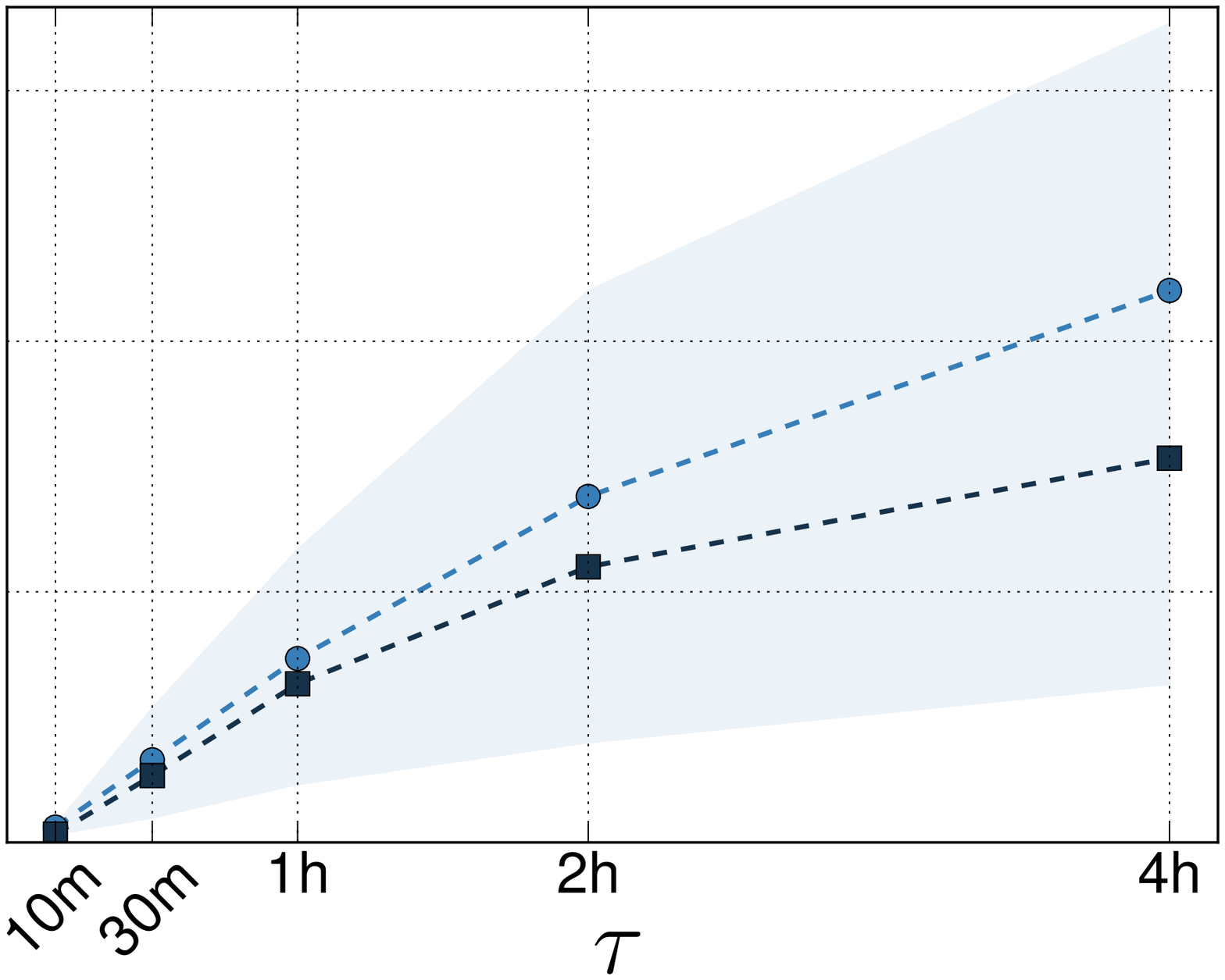}
}
\hspace*{-10.7pt}
\subfloat[\texttt{shn}]{\label{fig:tim_gran_vs_t_shn}
   \includegraphics[width=0.0624\textwidth]{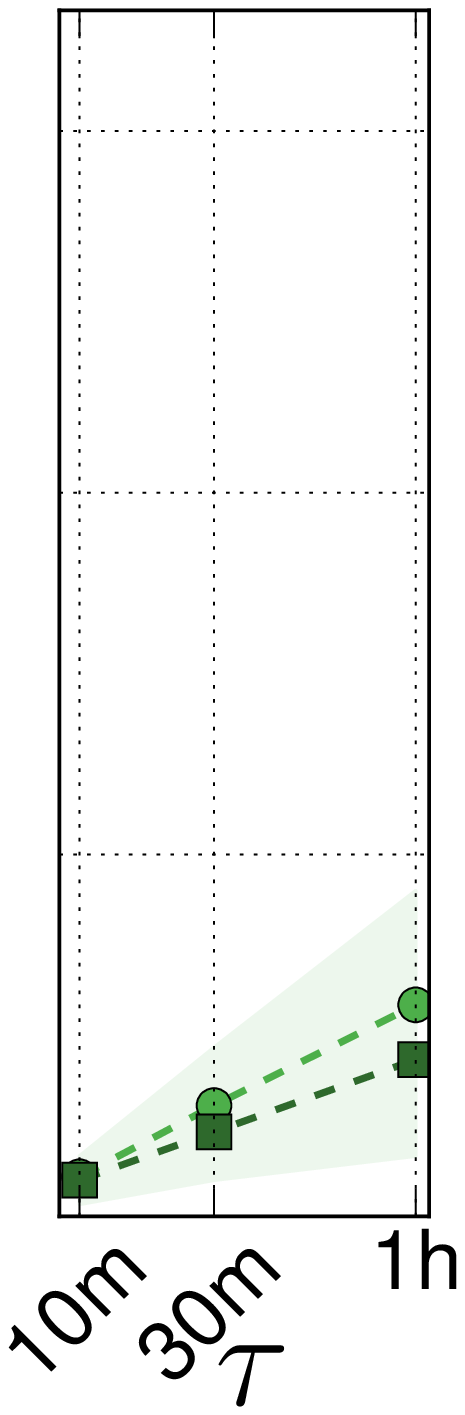}
}
%\vspace*{-4pt}
\caption{Spatial (a,b,c) and temporal (d,e,f) granularity versus the adversary knowledge $\tau$
in the citywide reference datasets.}
\label{fig:gran_vs_t}
%\vspace*{-7pt}
\end{figure*}

\begin{figure*}[tb]
\captionsetup[subfloat]{captionskip=-3pt}
\centering
\begin{minipage}[b]{0.68\textwidth}
	\centering
	%\hspace*{-5pt}
	\subfloat[\texttt{civ}]{\label{fig:pos_gran_vs_t_civ}
	   \includegraphics[height=0.265\textwidth]{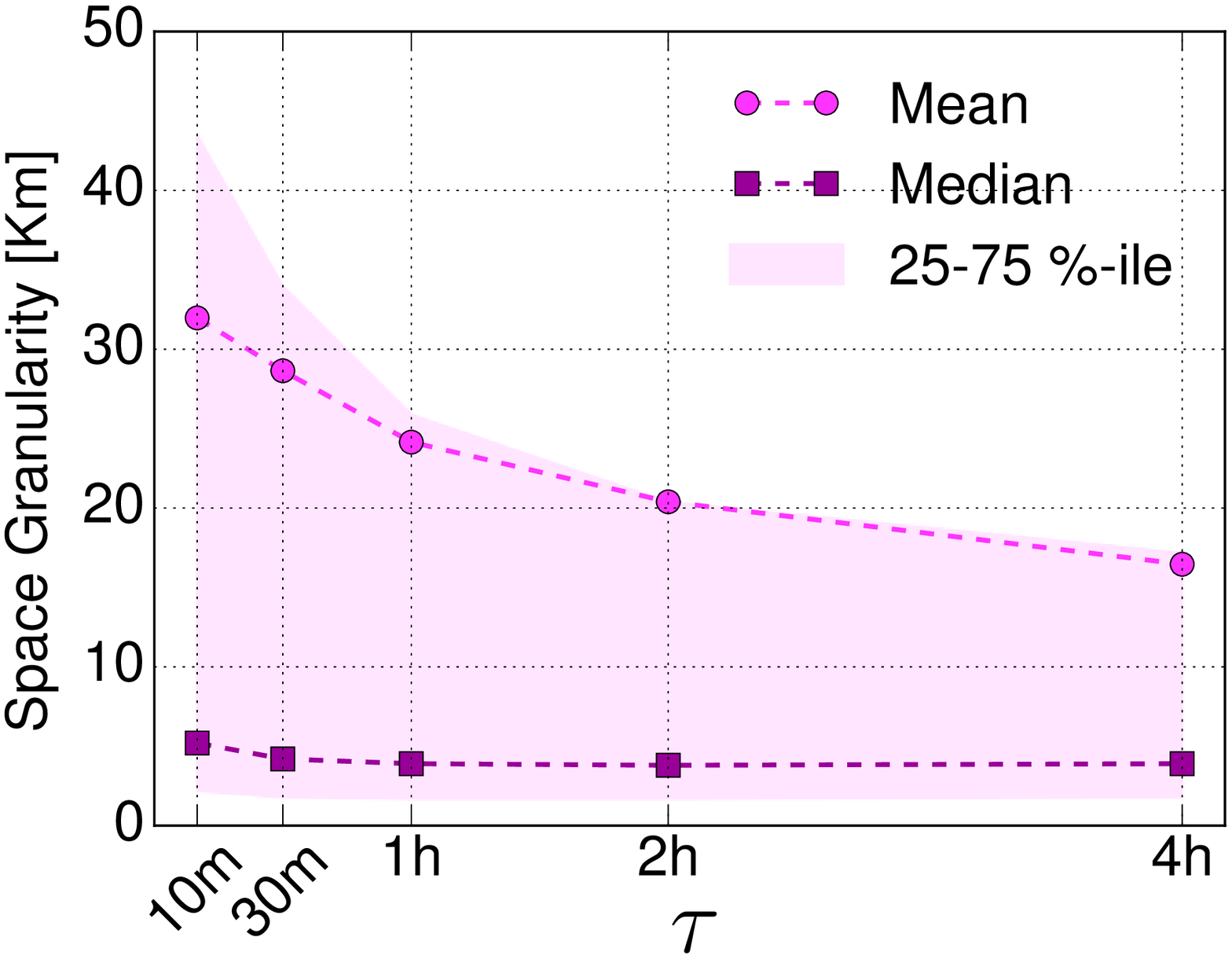}
	}
	\hspace*{-8pt}
	\subfloat[\texttt{sen}]{\label{fig:pos_gran_vs_t_sen}
	   \includegraphics[height=0.26\textwidth]{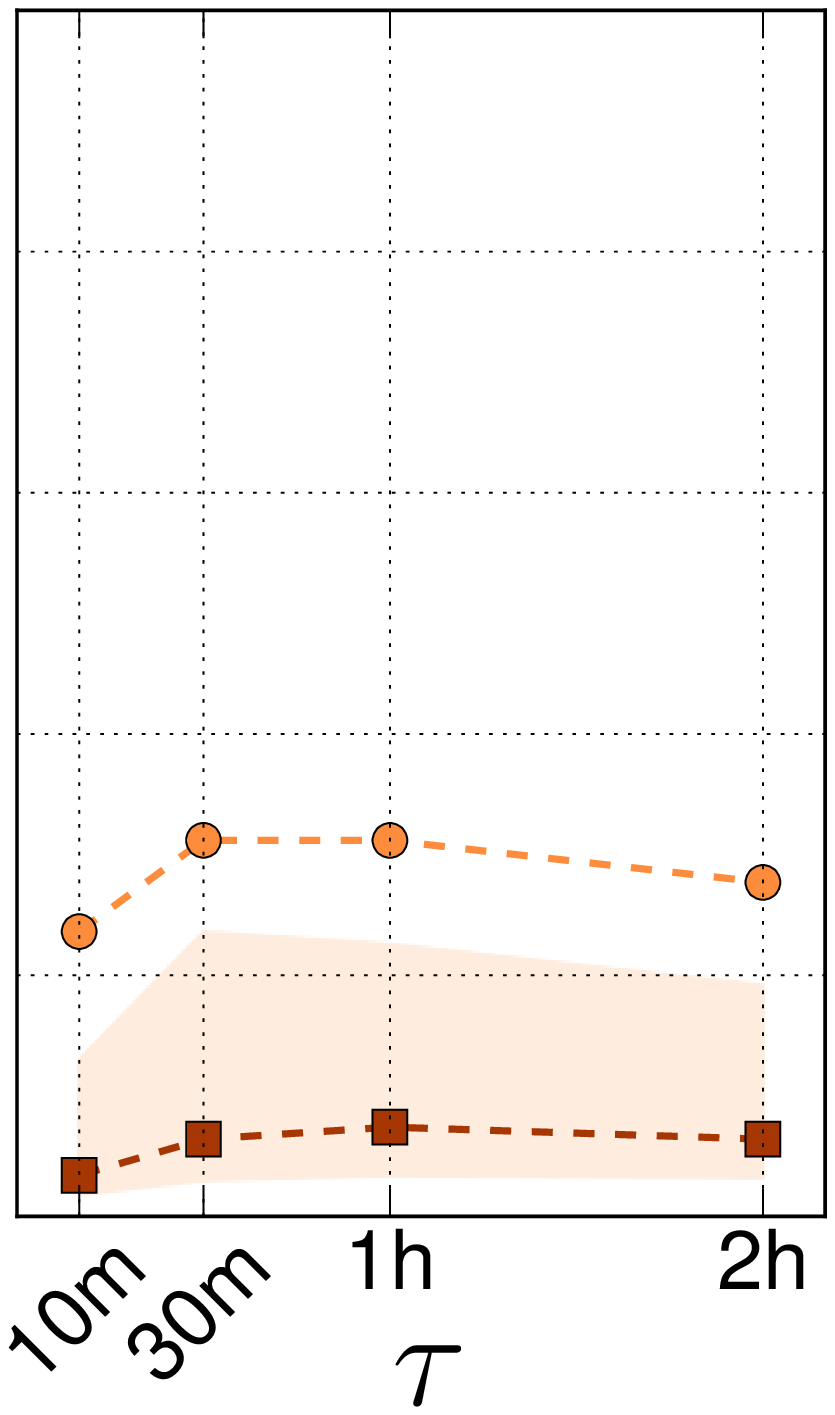}
	}
	\hspace*{-3pt}
	\subfloat[\texttt{civ}]{\label{fig:tim_gran_vs_t_civ}
	   \includegraphics[height=0.265\textwidth]{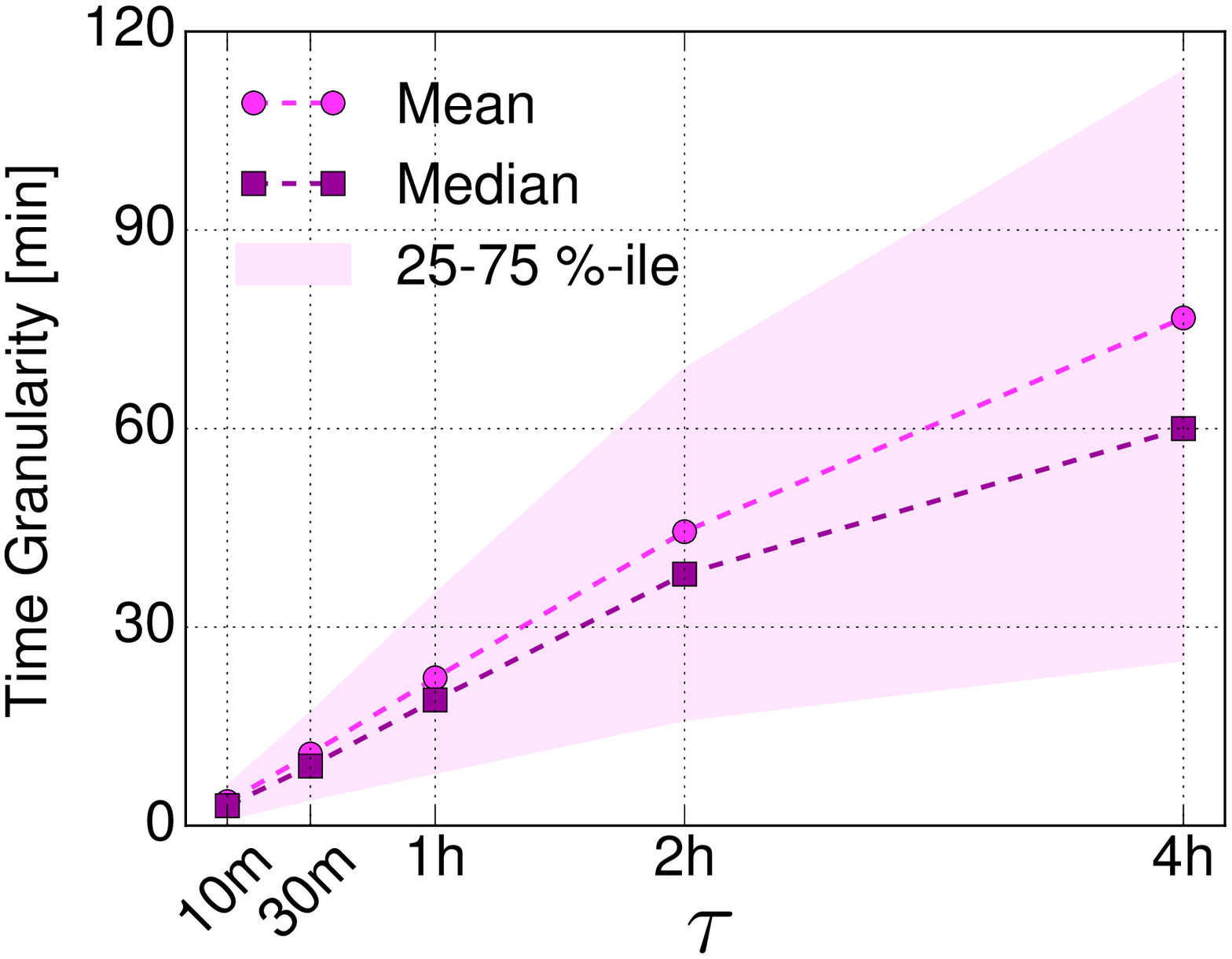}
	}
	\hspace*{-8pt}
	\subfloat[\texttt{sen}]{\label{fig:tim_gran_vs_t_sen}
	   \includegraphics[height=0.26\textwidth]{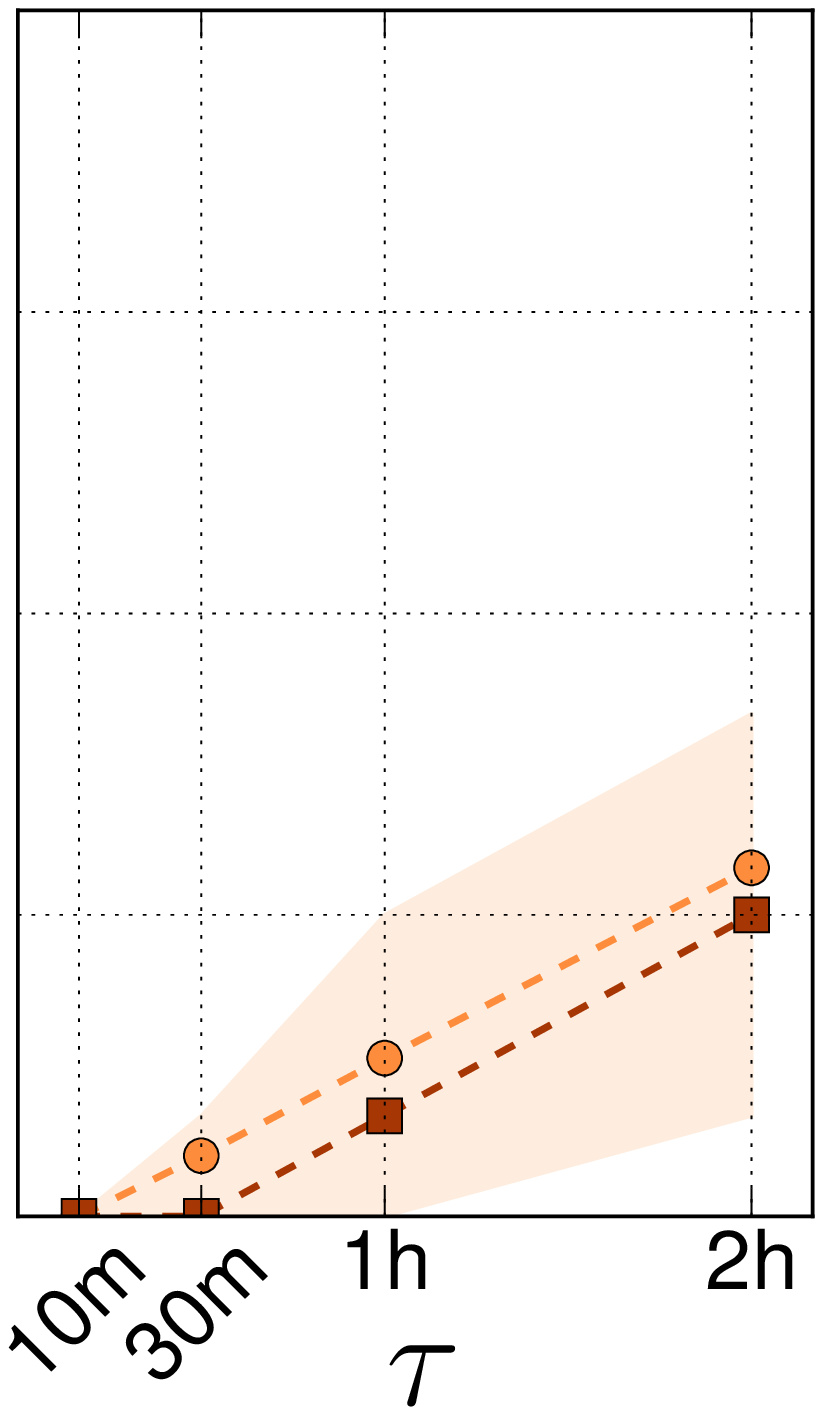}
	}
	\vspace*{-4pt}
	\caption{Spatial (a,b) and temporal (c,d) granularity versus $\tau$ % the adversary knowledge $\tau$
	in the nationwide reference datasets.}
	\label{fig:gran_vs_t_nation}
\end{minipage}
\hfill%\hspace*{10pt}
\begin{minipage}[b]{0.27\textwidth}
	\centering
	\includegraphics[width=0.82\textwidth]{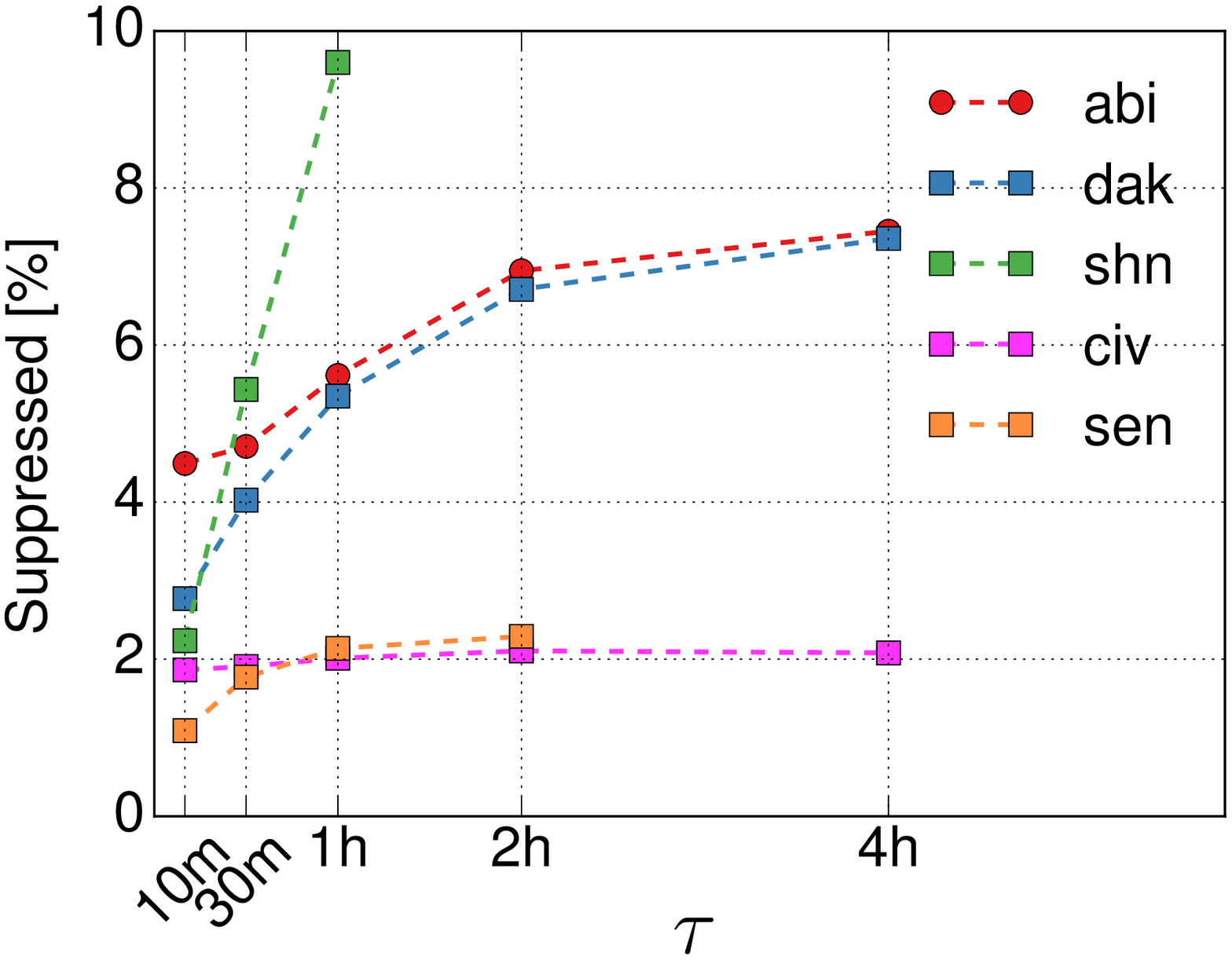}
	%\vspace*{-5pt}
	\caption{Suppressed samples versus $\tau$.} %the adversary knowledge $\tau$.}
	\label{fig:suppr_samples}
\end{minipage}
%\begin{minipage}[b]{0.3\textwidth}
%    \centering
%    \flushright
%    \renewcommand{\arraystretch}{1}
%    \setlength{\tabcolsep}{1.5pt}
%    \begin{tabular}{c}
%    %\subfloat[Space]{\label{fig:pos_gran_trend_abi}
%        \includegraphics[width=0.9\textwidth]{figures/d4d13_abi_pos_gran_trend.eps} \\
%    %}
%    %\subfloat[Time]{\label{fig:time_gran_trend_abi}
%        \includegraphics[width=0.9\textwidth]{figures/d4d13_abi_time_gran_trend.eps}\\
%    %}
%    \end{tabular}
%    \vspace*{-4pt}
%    \caption{Granularity time series, \texttt{abi} dataset.}
%    \label{fig:abi_trend}
%\end{minipage}
\vspace*{-10pt}
\end{figure*}

\subsection{Performance evaluation of \algokte}
\label{sub:peva-algokte}

We run \algokte on our reference datasets of mobile subscriber trajectories,
so that they are \anonymized. As the anonymized data are robust to probabilistic
attacks by design, we focus our evaluation on the cost of the anonymization, i.e.,
the loss of granularity. All results refer to the case of $2^{\tau,\epsilon}$-anonymization,
with $\epsilon=\tau$.

\subsubsection{Citywide datasets}

Fig.\,\ref{fig:gran_vs_t} portrays the mean,
median and first/third quartiles of the sample granularity in the \anonymized
citywide datasets \texttt{abi}, \texttt{dak} and \texttt{shn}. The plots show
how results vary when the adversary knowledge $\tau$ ranges from 10 minutes to 4 hours%
\footnote{The limited temporal span of the \texttt{shn} data prevents us from
testing attacks with knowledge $\tau$ higher than one hour. Indeed, a $\tau$
too close to the full dataset duration implies that the opponent has an a-priori
knowledge of the victim's trajectory that is comparable to that contained in the
data, making attempts at countering a probabilistic attack futile.}.
%
%\footnote{Probabilistic attacks are meaningless in presence of an adversary
%knowledge that is too close (e.g., around the same order of magnitude) of the
%full duration of trajectories in the dataset.
%Indeed, scuh a knowledgeable opponent would have no interest in running probabilistic
%attacks, since he already possesses the information the attack aims at revealing.
%Therefore, the limited temporal span of \texttt{uni-shn} makes it only sensible
%assessing \algokte performance in presence of an adversary knowledge $\tau$
%that is not larger than one hour.}.
They refer to the anonymized data granularity in space%
\footnote{The spatial granularity in Fig.\,\ref{fig:gran_vs_t} is expressed
as the sum of spans along the Cartesian axes. For instance, 1 km maps to, e.g., a
square of side 500 m.}, in Fig.\ref{fig:pos_gran_vs_t_abi}-\subref*{fig:pos_gran_vs_t_shn}
and time, in Fig.\ref{fig:tim_gran_vs_t_abi}-\subref*{fig:tim_gran_vs_t_shn}.

We remark how the \anonymized datasets retain significant levels of accuracy,
with a median granularity in the order of 1-3 km in space and below 45 minutes
in time. These levels of precision are largely sufficient for most analyses on
mobile subscriber activities, as discussed in, e.g.,~\cite{coscia12}.
The temporal granularity is negatively affected by an increasing adversary
knowledge $\tau$, which is expected. Interestingly, however, the spatial
granularity is only marginally impacted by $\tau$: protecting the data from a
more knowledgeable attacker does not have a significant cost in terms of
spatial accuracy.

\subsubsection{Nationwide datasets}

Fig.\,\ref{fig:gran_vs_t_nation} shows equivalent results for the nationwide
datasets \texttt{civ} and \texttt{sen}. The evolution of temporal granularity
versus $\tau$, in Fig.\ref{fig:tim_gran_vs_t_civ}-\subref*{fig:tim_gran_vs_t_sen}
is consistent with citywide scenarios.
Differences emerge in terms of spatial granularity: in the \texttt{civ} case
(Fig.\ref{fig:pos_gran_vs_t_civ}) a reversed trend emerges, as accuracy grows
along with the attacker knowledge.
This counterintuitive result is explained by the thin user presence in the
\texttt{civ} dataset: as per Tab.\,\ref{tbl:dataset_stats}, \texttt{civ} has
a density of subscribers per Km$^2$ that is one or two orders of magnitude
lower than those in our other reference datasets.
Such a geographical sparsity makes it difficult to find individuals with
similar spatial trajectories: increasing $\tau$ has then the effect of enlarging
the set of candidate trajectories for merging at each epoch,
%In turn, this leads to a higher probability to be presented
%with users whose sub-trajectories are similar, and thus to an improvement
%of accuracy in the generalized data.
with a positive influence on the accuracy in the generalized data.

These considerations are confirmed by the results with the
\texttt{sen} dataset (Fig.\ref{fig:pos_gran_vs_t_sen}). As per
Tab.\,\ref{tbl:dataset_stats}, this dataset
features a subscriber density that is about one order of magnitude higher
than that of \texttt{civ}, but around one order of magnitude lower than those
of the \texttt{abi}, \texttt{dak} and \texttt{shn}. Coherently, the
spatial granularity trend falls in between those observed for
such datasets, and it is not positively or negatively impacted by the
attacker knowledge.

More generally, the results in Fig.\,\ref{fig:gran_vs_t_nation}
demonstrate that \algokte can scale to large-scale real-world datasets.
The absolute performance is good, as the \anonymized data retains
substantial precision: the median levels of granularity in space
and time are comparable to those achieved in citywide datasets.
Finally, we remark that, in all cases, the amount of samples suppressed by
\algokte is in the 1\%--7\% range.

\subsubsection{Sample suppression}

The amount of samples suppressed by \algokte in the \anonymization process
is portrayed in Fig.\,\ref{fig:suppr_samples}. We note that resorting to
suppression becomes more frequent as the adversary knowledge increases.
However, even when the opponent is capable of tracking a user during
four continued hours, the percentage of suppressed samples remains low,
typically well below 10\%. Moreover, the trend in the long-timespan datasets
is clearly sublinear, suggesting that suppression does not become prevalent
with higher $\tau$. Results are fairly consistent across citywide datasets%
\footnote{The spurious point at $\tau$ = 1 hour in \texttt{shn} is due to the
fact that the time interval $\tau+\epsilon$ is already very large, at around
the same order of magnitude of the full dataset duration.}. Nationwide datasets
are also aligned, and yield even lower suppression rates, at around 2\%.
This difference is explained by the fact that a larger number of users
allows for a more efficient spectral clustering in \algokte.

\subsubsection{Disaggregation over time}

\begin{figure*}[tb]
\centering
\subfloat[\texttt{abi}, space]{\label{fig:pos_gran_trend_abi}
   \includegraphics[width=0.32\textwidth]{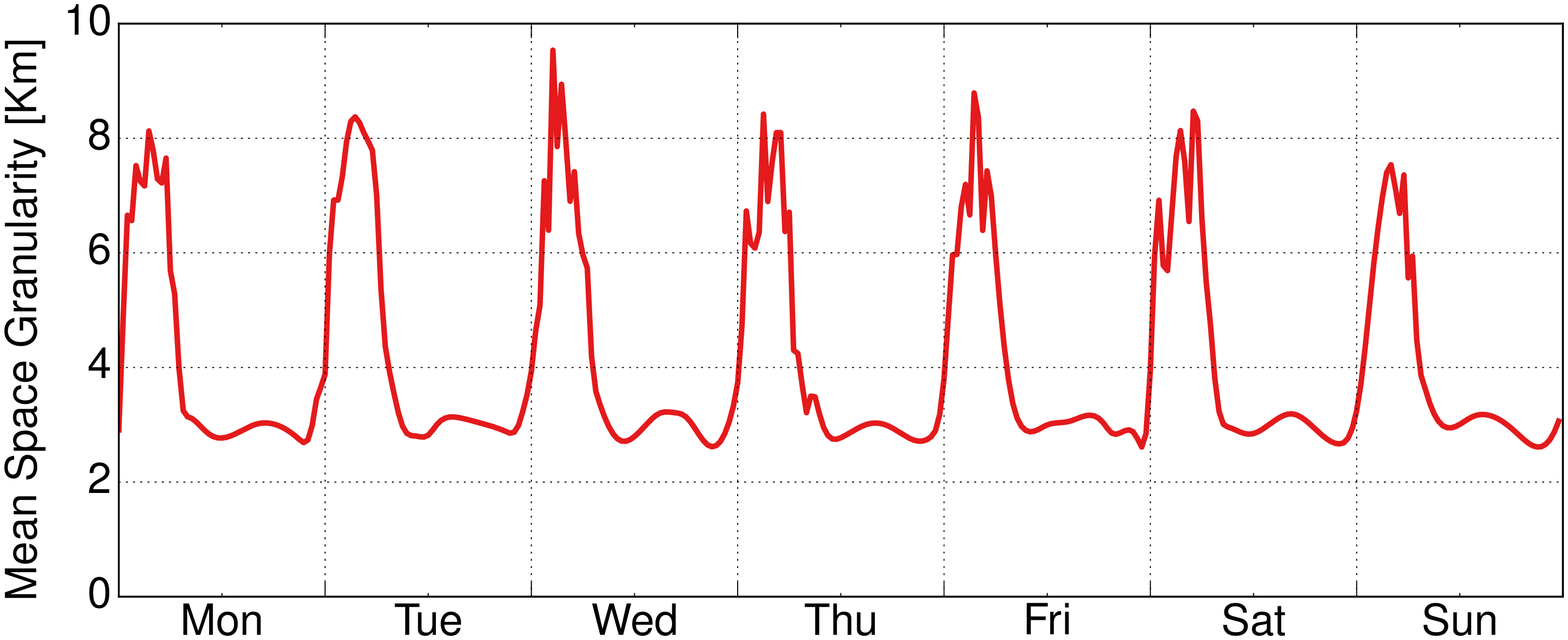}
}
\subfloat[\texttt{abi}, time]{\label{fig:time_gran_trend_abi}
   \includegraphics[width=0.32\textwidth]{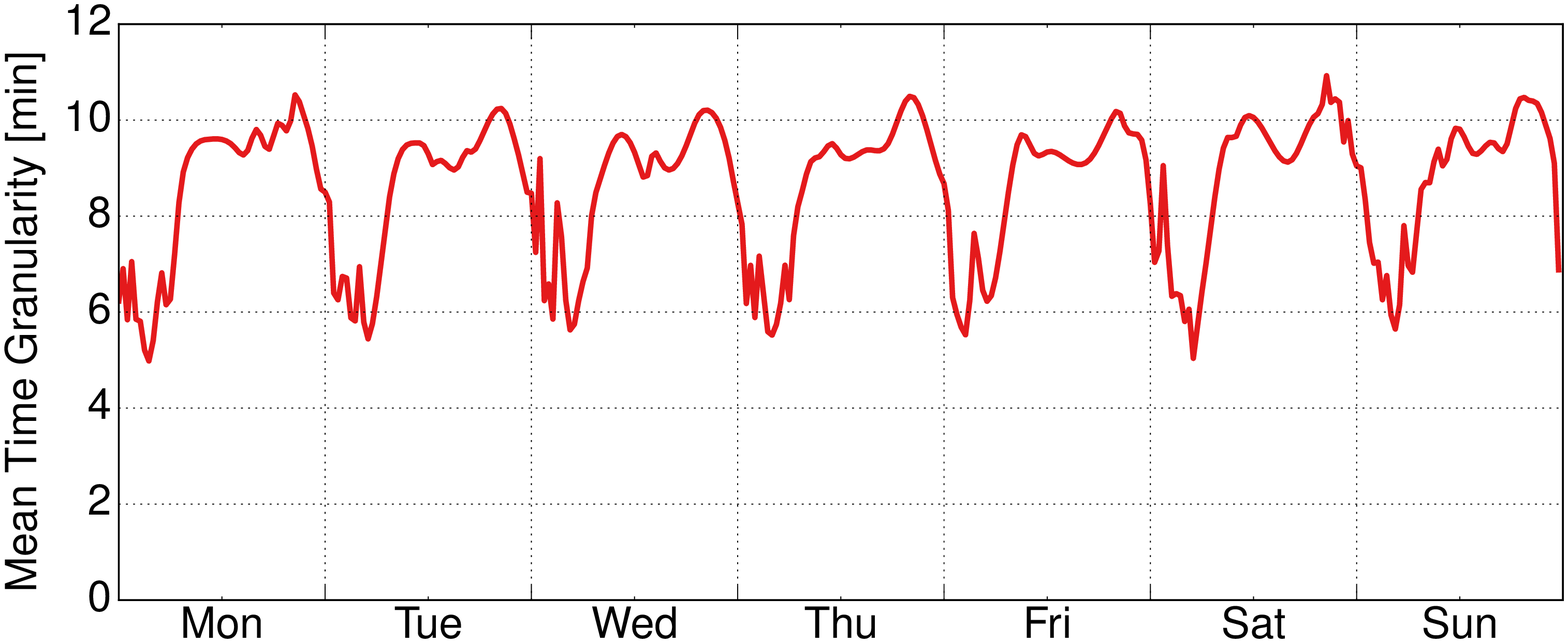}
}
\subfloat[\texttt{abi}, suppression]{\label{fig:blanked_trend_abi}
   \includegraphics[width=0.32\textwidth]{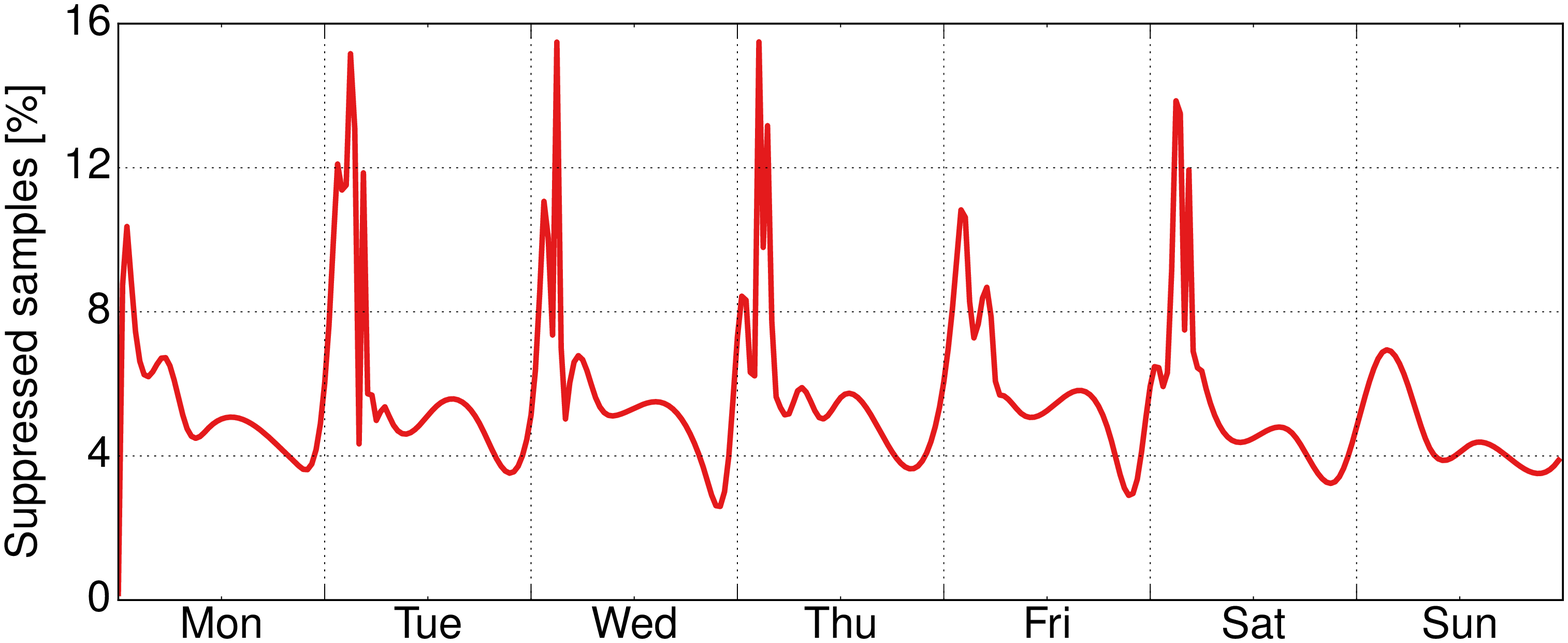}
}
%\vspace*{-3pt}
\caption{Time series of spatiotemporal accuracy (a,b) and suppression usage (c)
for one sample week in the \texttt{abi} dataset.}
\label{fig:abi_trend}
%\vspace*{-4pt}
\end{figure*}

As an intriguing concluding remark, Fig.\,\ref{fig:abi_trend} reveals a clear
circadian rhythm in the granularity of \anonymized data, as well as in the
percentage of suppressed samples. The plots refer to one sample week in the
\texttt{abi} and \texttt{dak} datasets, when $\tau$ = 30 min, but consistent
results were observed in all of our reference datasets.
Specifically, the mean spatial granularity, in Fig.\,\ref{fig:pos_gran_trend_abi},
is much finer during daytime, when subscribers are more active and the volume of
trajectories is larger: here, it is easier to hide a user into the crowd.
Overnight displacements are instead harder to anonymize, since subscribers
are limited in number and they tend to have diverse patterns.
This is also corroborated by the significantly higher suppression of samples
between midnight and early morning, in Fig.\,\ref{fig:blanked_trend_abi}.
Time granularity, in Fig.\,\ref{fig:time_gran_trend_abi}, is less subject to
day-night oscillations: the slightly higher accuracy recorded at night
is an artifact of the important relative suppression of samples at those times.

\subsubsection{Summary}

Overall, our results show that \algokte attains \anon of real-world datasets
of mobile traffic, while maintaining a remarkable level of accuracy in the data.
Interestingly, its performance is better when most needed, at daytime, when
the majority of human activities take place.

\section{Related work}
\label{sec:related}

Protection of individual mobility data has attracted significant
attention in the past decade. However, attack models and privacy
criteria are very specific to the different data collection contexts.
Hence, solutions developed for a specific type of movement data are
typically not reusable in other environments.

For instance, a vast amount of works have targeted user privacy in
location-based services (LBS). There, the goal is ensuring that
single georeferenced queries are not uniquely identifiable~\cite{gruteser03}. %,gedik08}.
This is equivalent to anonymizing each spatiotemporal sample independently,
and a whole other problem from protecting full trajectories.
Even when considering sequences of queries, the LBS milieu allows
pseudo-identifier replacement, and most solutions rely on this
approach, see, e.g.,~\cite{meyerowitz09,hoh07}.
If applied to spatiotemporal trajectories, these techniques would
seriously and irreversibly break up trajectories in time, disrupting
data utility.

Another popular context is that of spatial trajectories that do not
have a temporal dimension. The problem of anonymizing datasets of
spatial trajectories has been thoroughly explored in data mining,
and many practical solutions based on generalization have been
proposed, see, e.g.,~\cite{nergiz09,monreale10,chen12,poulis14}.
Such solutions are not compatible with or easily extended
to the more complex spatiotemporal data we consider.

Some works explicitly target privacy preservation of spatiotemporal
trajectories. However, the precise context they refer to makes again
all the difference.
First, most such solutions consider scenarios where user movements
are sampled at regular time intervals that are identical for all
individuals~\cite{abul08,domingo-ferrer12}, % yarovoy09,
or where the
number of samples per device is very small~\cite{fung09}.
These assumptions hold, e.g., for GPS logs or RFID record, but
not for trajectories recorded by mobile operators: the latter are
irregularly sampled, temporally sparse, and cover long time periods,
which results in at least hundreds of samples per user.
Second, many of the approaches above disrupt data utility, by, e.g.,
trimming trajectories~\cite{song14}, or violate the principles of
PPDP, by, e.g., perturbating or permutating the
trajectories~\cite{abul08,domingo-ferrer12},
or creating fictitious samples~\cite{abul10}.
Third, all previous studies aim at attaining $k$-anonymity of
spatiotemporal trajectories, i.e., they protect the data against
record linkage; this includes recent work specifically tailored
to mobile subscriber trajectory datasets~\cite{gramaglia15}.
As explained in Sec.\,\ref{sec:reqs}, $k$-anonymity is only a
partial countermeasure to attacks on spatiotemporal trajectories.

Provable privacy guarantees are instead offered by {\it differential privacy},
which commends that the presence of a user's data in the published dataset
should not change substantially the output of the analysis, and thus formally
bounds the privacy risk of that user~\cite{dwork06}.
There have been attempts at using differential privacy with mobility data.
Specifically, it has been successfully used the in the LBS context, when
publishing aggregate information about the location of a large number of
users, see, e.g.,~\cite{chen13}. However, the requirements of these solutions
already become too strong in the case of individual LBS access data~\cite{chatzikokolakis14}.
To address this problem, a variant of differential privacy, named
{\it geo-indistinguishability} has been introduced: it requires that any
two locations become more indistinguishable as they are geographically
closer~\cite{andres13}. Practical mechanisms achieve geo-indistinguishability,
see, e.g.,~\cite{chatzikokolakis14,andres13}.
However, all refer to the anonymization of single LBS queries:
as of today, differential privacy and its derived definitions still
appear impractical in the context of spatiotemporal trajectories.

%%% --- EXTENSION
%To the best of our knowledge, there exist no solution to probabilistic
%attacks on spatiotemporal trajectories. As discussed in Sec.\,\ref{sub:att},
%this type of attacks is highly relevant in the context of mobile user
%trajectory data, and it is more challenging to address than record linkage.
%In fact, countering probabilistic attacks requires a privacy model that
%extends standard $k$-anonymity; anonymization approaches that implement
%such a model implicitly offer protection against both probabilistic and
%record linkage attacks.
%This allowed us to show the performance gain provided by a subset of our
%solution over current state-of-the-art techniques for the $k$-anonymization
%of spatiotemporal trajectories, i.e., W4M~\cite{abul10} and GLOVE~\cite{gramaglia15}.

%Trujillo criticizes $(k,\delta)$-anonymity in NWA\cite{abul08} and W4M~\cite{abul10}.

%\section*{Acknowledgments}

\section{Conclusions}
\label{sec:conclusion}

In this paper, we presented a first PPDP solution to probabilistic
and record linkage attacks against mobile subscriber trajectory data.
To that end, we introduced a novel privacy model, \anon, which generalizes
the popular criterion of $k$-anonymity.
Our proposed algorithm, \algokte, implements \anon in real-world
datasets, while retaining substantial spatiotemporal accuracy in the
anoymized data.
%%% --- EXTENSION
%In the future, we plan to extend our solutions so as to cope with
%probabilistic attacks under even more generic attacker knowledge,
%i.e., disjoint time intervals: this currently remains an open
%problem.



%\bibliographystyle{IEEEtran}

\end{document}